\documentclass[twocolumn,showpacs,preprintnumbers,amsmath,amssymb,superscriptaddress,aps,prl]{revtex4-1}
\usepackage[maccyr]{inputenc}
\usepackage[T2A]{fontenc}
\usepackage[english, russian]{babel}
\usepackage{bm}
\usepackage{epsfig,amssymb,amsmath,bm}
\usepackage[usenames,dvipsnames,svgnames,table]{xcolor}

\begin{document}

\title{Hypercomplex representation of the Lorentz's group}
\author{K. S. Karplyuk}
\email{karpks@hotmail.com}
 \affiliation{Taras Shevchenko University, Academic
Glushkov prospect 2, building 5, Kyiv 03122, Ukraine, karpks@hotmail.com}
\author{O. O. Zhmudskyy}\email{ozhmudsk@ucf.edu}
 \affiliation{University of Central Florida, 4000 Central Florida Blvd. Orlando, FL, 32816, ozhmudsk@ucf.edu}
\begin{abstract}
Lorentz's group represented by the hypercomplex system of numbers, which is based on Dirac's matrices, is investigated.
This representation is similar to the space rotation representation by quaternions.  This representation has several advantages.
Firstly, this is a reducible representation. That is why transformation of different geometrical objects (vectors, antisymmetric tensors
of the second order and bispinors) is implemented by the same operators.  Secondly, the rule of composition of two
arbitrary Lorentz's transformations has a simple form.  These advantages strongly simplify finding many of the laws related to the
Lorentz's group.  In particular, they simplify investigation of the spin connection with the Pauli-Lubanski pseudovector and the Wigner little group.
\end{abstract}

\pacs{12., 12.20.-m, 13.66.-a}

\maketitle
\section{Introduction}
The main properties of the Lorentz's group have been investigated using infinitesimal transformations.
However, finite transformations are necessary for many practical applications.
Several consistent transformations of this kind are often necessary.
In this case, transformation progress depends on simplicity/complexity of the composition law of the parameters, which
describe the transformation.   This, in turn, depends on the choice of the parameters.
The simplest composition rule is Fedorov's  
parametrization. It introduced matrix $L$, which reflects 
Lorentz's transformation of the 4D vector, in the form shown below.  \\    
I.e. it parametrized this L matrix by the complex 3D vector $\bm{c}=\bm{a}+i\bm{b}$\cite{f}.
\begin{widetext}
\[L(\bm{c})=2\left[
\begin{array}{cccc}
a_1a_1+b_1b_1+\frac{1-a^2-b^2}{2}&a_1a_2+b_1b_2-a_3&a_1a_3+b_1b_3+a_2&i(b_1+\varepsilon_{1kl}a_kb_l)\\
a_2a_1+b_2b_1+a_3&a_2a_2+b_2b_2+\frac{1-a^2-b^2}{2}&a_2a_3+b_2b_3-a_1&i(b_2+\varepsilon_{2kl}a_kb_l)\\
a_3a_1+b_3b_1-a_2&a_3a_2+b_3b_2+a_1&a_3a_3+b_3b_3+\frac{1-a^2-b^2}{2}&i(b_3+\varepsilon_{3kl}a_kb_l)\\
-i(b_1-\varepsilon_{1kl}a_kb_l)&-i(b_2-\varepsilon_{2kl}a_kb_l)&-i(b_3-\varepsilon_{3kl}a_kb_l)&\frac{1+a^2+b^2}{2}
\end{array}
\right]\]
\end{widetext}  
In this way Fedorov shows that two successive transformations  $L(\bm{c}_1)$ and $L(\bm{c}_2)$ with the parameters
  $\bm{c}_1$ and $\bm{c}_2$ can be changed by the single one $L(\bm{c})=L(\bm{c}_2)L(\bm{c}_1)$.  Where parameter  $\bm{c}$ is expressed  by
   $\bm{c}_1$ и $\bm{c}_2$ as
\begin{equation}
\bm{c}=\frac{\bm{c}_1+\bm{c}_2+\bm{c}_2\times\bm{c}_1}{1-\bm{c}_1\cdot\bm{c}_2}.
\end{equation}
However, the use of matrix  $L(\bm{c})$ in relativistic quantum electrodynamics is inconvenient and is not widely used.
Instead matrix exponent $e^{L_{\alpha\beta}\sigma^{\alpha\beta}}$ for Dirac's bispinor transformation  is widely used. 
In this case, the transformation parameter is the antisymmetric tensor $L_{\alpha\beta}$ of the second order presented in 
the matrix form  $L_{\alpha\beta}\sigma^{\alpha\beta}$.   In fact, this representation is the hypercomplex number form.
However, the composition rule for these hypercomplex parameters for the two arbitrary Lorentz's transformations is not known.
We will get this rule below and show that the composition rule for hypercomplex parameters is similar to the Fedorov's (1).
But Lorentz's transformations are implemented in a way that is traditional for relativistic quantum electrodynamics. \\
It will also be shown that hypercomplex description allows to easily derive the set of rules connected with Lorentz's transformations.
\section{Dirac's hypercomplex numbers}
It is known, that unit matrix $\hat{1}$, four Dirac's matrices $\gamma^\alpha$
\begin{equation}
\gamma^\alpha\gamma^\beta+\gamma^\beta\gamma^\alpha=2\eta^{\alpha\beta}\hat{1},
\end{equation}
$\eta^{\alpha\beta}=\rm{diag}(1,-1,-1,-1)$, and eleven products
\begin{equation}
\hat{\iota}=\gamma^0\gamma^1\gamma^2\gamma^3,\hspace{2mm}\pi^\alpha=\gamma^\alpha\hat{\iota},\hspace{2mm} \sigma^{\alpha\beta}=(\gamma^\alpha\gamma^\beta-\gamma^\beta\gamma^\alpha)/2
\end{equation}
can be used as sixteen basic units of the hypercomplex numbers system \cite{b,s}.  It does not matter what the actual
presentation is. Frequently used are: standard Dirac-Pauli basis and chiral Weyl's basis.

Let us call numbers of this system
\begin{equation}
D=a\hat{1}+b \hat{\iota}+c_\alpha\gamma^\alpha+d_\alpha\pi^\alpha+f_{\alpha\beta}\sigma^{\alpha\beta}
\end{equation}
 Dirac's numbers.
Coefficients in front of the basic units are complex numbers.
Assume that coefficients $f_{\alpha\beta}$ in front of $\sigma^{\alpha\beta}$ are antisymmetric $f_{\alpha\beta}=-f_{\beta\alpha}$.  It is possible
because their symmetrical parts have no contribution into the contraction with the antisymmetric $\sigma^{\alpha\beta}$. As usual, the Greek indices take on values 0,1,2,3, 
Latin indices - 1,2,3.  The hypercomplex representation of the continuous proper Lorentz's transformations are discussed below.  
It will be shown that numbers $a\hat{1}$ and $a\hat{\iota}$ under the continuous proper Lorentz's transformations behave as scalars $a_\alpha\gamma^\alpha$, $a_\alpha\pi^\alpha$  as vectors,  and $a_{\alpha\beta}\sigma^{\alpha\beta}$  as antisymmetric tensors of the second order.
But under the discrete space reflection (inversion) like $\gamma^0D\gamma^0$, numbers
 $a\hat{1}$ behave as scalars, $\gamma^0a\hat{1}\gamma^0=a\hat{1}$,  numbers  $a\hat{\iota}$ as pseudoscalars, $\gamma^0a\hat{\iota}\gamma^0=-a\hat{\iota}$, numbers $a_\alpha\gamma^\alpha$ as vectors $\gamma^0(a_0\gamma^0+a_k\gamma^k)\gamma^0=a_0\gamma^0-a_k\gamma^k$, numbers  $a_\alpha\pi^\alpha$ as pseudovectors $\gamma^0(a_0\pi^0+a_k\pi^k)\gamma^0=-a_0\pi^0+a_k\pi$.  This is why we will use corresponding terms for these variables.

We can add, subtract and multiply Dirac's numbers as any matrices.  Table 1 can be useful in order to simplify multiplications.
First, the multiplier must be taken from the left column of the table and the second one from the first row.  For example,
\begin{gather}
a_{\alpha\beta}\sigma^{\alpha\beta}b_\mu\pi^\mu=2b^\alpha(a^\diamond_{\alpha\beta}\gamma^\beta-a_{\alpha\beta}\pi^\beta).
\end{gather}
Here $a^\diamond_{\alpha\beta}$ is a tensor dual to the  $a_{\alpha\beta}$ one
\begin{gather}
a^\diamond_{\alpha\beta}=\frac{1}{2}\varepsilon_{\alpha\beta\mu\nu} a^{\mu\nu},
\end{gather}
$\varepsilon^{\alpha\beta\mu\nu}$ -- completely antisymmetric  tensor, $\varepsilon^{0123}=1$, $\varepsilon_{0123}=-1$.
\begin{widetext}
\begin{center}{Table 1}\\
\begin{tabular}{|c|c|c|c|c|c|}
\hline
&$b_\mu\gamma^\mu$&$b_\mu\pi^\mu$&$b_{\mu\nu}\sigma^{\mu\nu}$&$b\hat{\iota}$
\\ \hline
$a_\alpha\gamma^\alpha$&\parbox[c]{3cm}{\begin{center}$a_\alpha b^\alpha+\frac{1}{2}[a_\alpha b_\beta]\sigma^{\alpha\beta}$\end{center}}
&\parbox[c]{3.3cm}{\begin{center}$\hat{\iota}a_\alpha b^\alpha+\frac{1}{2}[a_\alpha b_\beta]^\diamond\sigma^{\alpha\beta}$\end{center}}
&\parbox[c]{3.5cm}{\begin{center}$2(a^\alpha b_{\alpha\beta}\gamma^\beta-a^\alpha b^\diamond_{\alpha\beta}\pi^\beta)$\end{center}}
&$ba_\alpha\pi^\alpha$
\\ \hline
$a_\alpha\pi^\alpha$&\parbox[c]{3.5cm}{\begin{center}$-\hat{\iota}a_\alpha b^\alpha-
\frac{1}{2}[a_\alpha b_\beta]^\diamond\sigma^{\alpha\beta}$\end{center}}
&\parbox[c]{3cm}{\begin{center}$a_\alpha b^\alpha+\frac{1}{2}[a_\alpha b_\beta]\sigma^{\alpha\beta}$\end{center}}
&\parbox[c]{3.5cm}{\begin{center}$2(a^\alpha b^\diamond_{\alpha\beta}\gamma^\beta+a^\alpha b_{\alpha\beta}\pi^\beta)$\end{center}}
&$-ba_\alpha\gamma^\alpha$
\\ \hline
$a_{\alpha\beta}\sigma^{\alpha\beta}$&\parbox[c]{4cm}{\begin{center}$-2(b^\alpha a_{\alpha\beta}\gamma^\beta+b^\alpha a^\diamond_{\alpha\beta}\pi^\beta)$
\end{center}}
&\parbox[c]{4cm}{\begin{center}$2(b^\alpha a^\diamond_{\alpha\beta}\gamma^\beta-b^\alpha a_{\alpha\beta}\pi^\beta)$\end{center}}
&\parbox[c]{4cm}{\begin{center}
$-2a^{\alpha\beta}\!\!\bullet b_{\alpha\beta}+2[a_{\alpha\mu}b_{.\beta}^{\mu.}]\sigma^{\alpha\beta}$
\end{center}}
&$ba^\diamond_{\alpha\beta}\sigma^{\alpha\beta}$
\\ \hline
$a\hat{\iota}$&$-ab_\mu\pi^\mu$&$ab_\mu\gamma^\mu$&$a b^\diamond_{\alpha\beta}\sigma^{\alpha\beta}$&$-ab$\\
\hline
\end{tabular}
\end{center}
\end{widetext}
In Table 1 unit vector  $\hat{1}$ is omitted. Below we also omit $\hat{1}$  if it does not cause misunderstanding.

 Square brackets designate two operations.  First, the antisymmetrized tensor product that  collates 
 two 4D vectors to the antisymmetric tensor of the second order
\begin{gather}
[a_\alpha b_\beta]=a_\alpha b_\beta-b_\alpha a_\beta,\hspace{7mm} [a_\alpha b_\beta]=-[a_\beta b_\alpha].
\end{gather}
 Second, the antisymmetrized contraction by one index that collates 
 the two  antisymmetric tensors of the second order a similar
  antisymmetric tensor 
\begin{gather}
[a_{\alpha\mu} b_{.\beta}^{\mu.}]=a_{\alpha\mu} b_{.\beta}^{\mu.}-b_{\alpha\mu} a_{.\beta}^{\mu.},\hspace{2mm}
[a_{\alpha\mu} b_{.\beta}^{\mu.}]=-[a_{\beta\mu} b_{.\alpha}^{\mu.}].
\end{gather}
Symbol  \lq\lq bullet\rq\rq\, ($\bullet$) designates the operation that collates two  antisymmetric tensor to the scalar and
pseudoscalar:
\begin{gather}
a^{\alpha\beta}\!\!\bullet b_{\alpha\beta}=a^{\alpha\beta}b_{\alpha\beta}-\hat{\iota}a^{\alpha\beta}b^\diamond_{\alpha\beta}=\nonumber\\
=a^{\alpha\beta}b_{\alpha\beta}-\frac{1}{2}\hat{\iota}\varepsilon^{\alpha\beta\mu\nu}a_{\alpha\beta}b_{\mu\nu}.
\end{gather}

The system of numbers (4) contains several subsystems.  Below we will use three of them.

Numbers $x=a\hat{1}+\hat{\iota}b$ form a subsystem, isomorphic to the complex numbers. Let us call these numbers $\hat{\iota}$-complex numbers.
Consequently, let us call $\breve{x}=a\hat{1}-\hat{\iota}b$ numbers as $\hat{\iota}$-conjugate to the  $x=a\hat{1}+\hat{\iota}b$ numbers.
We can use $\hat{\iota}$-complex numbers in the same way as complex numbers.  In particular
\begin{equation}
\hat{\iota}\hat{\iota}=-\hat{1},\hspace{7mm} \sqrt{-\hat{1}}=\pm\hat{\iota}.
\end{equation}
Below we will use $\hat{\iota}$-complex numbers and $\hat{\iota}$-complex vectors.

Four basic units  $\hat{1}$, $\sigma^{kl}$, where $k,l=1,2,3$ form another subsystem from the numbers:
\begin{equation}
q=a\hat{1}+f_{kl}\sigma^{kl}
\end{equation}
This subsystem is isomorphic to the quaternions. The multiplication table for the basic units $\hat{1}$, $\sigma^{23}$, $\sigma^{31}$, $\sigma^{12}$
coincides with the quaternion multiplication table:
\[\mbox{\begin{tabular}{|c|c|c|c|} \hline \hline
 {$\times$}&$\sigma^{23}$&$\sigma^{31}$&$\sigma^{12}$\\
 \hline $\sigma^{23}$&-$\hat{1}$&$\sigma^{12}$&$-\sigma^{31}$\\
 \hline $\sigma^{31}$&$-\sigma^{12}$&-$\hat{1}$&$\sigma^{23}$\\ \hline
        $\sigma^{12}$&$\sigma^{31}$&$-\sigma^{23}$&-$\hat{1}$\\ \hline \hline
\end{tabular}}\hspace{13mm}\mbox{\begin{tabular}{|c|c|c|c|} \hline \hline
 {$\times$}&$\bm{i}$&$\bm{j}$&$\bm{k}$\\
 \hline $\bm{i}$&-1&$\bm{k}$&$-\bm{j}$\\
 \hline $\bm{j}$&$-\bm{k}$&-1&$\bm{i}$\\ \hline
        $\bm{k}$&$\bm{j}$&$-\bm{i}$&-1\\ \hline \hline
\end{tabular} }\]
Also numbers
\begin{equation}
d=a\hat{1}+b \hat{\iota}+f_{\alpha\beta}\sigma^{\alpha\beta}
\end{equation}
form subsystem, which is based on 8 basic units $\hat{1}$, $\hat{\iota}$, $\sigma^{\alpha\beta}$.
Numbers $\hat{1}$ and $\hat{\iota}$ commute to each other and with all $\sigma^{\alpha\beta}$.
In this sense, numbers  $a\hat{1}+b\hat{\iota}$ from the subsystem (12) are similar to the usual complex numbers $a+bi$. But multiplication
by $\hat{\iota}$ converts  $\sigma^{lm}$ into $\sigma^{0k}$ because
\begin{gather}
\hat{\iota}\sigma^{\alpha\beta}=\frac{1}{2}\varepsilon^{\alpha\beta..}_{..\mu\nu}\sigma^{\mu\nu}.
\end{gather}
In particular, $\hat{\iota}\sigma^{lm}=\frac{1}{2}\varepsilon^{lm..}_{..\alpha\beta}\sigma^{\alpha\beta}$. For example,
\begin{equation}
\hat{\iota}\sigma^{23}=-\sigma^{01},\hspace{7mm}\hat{\iota}\sigma^{31}=-\sigma^{02},\hspace{7mm}\hat{\iota}\sigma^{12}=-\sigma^{03}.
\end{equation}
That is why instead of the subsystem (12) with 8 units, we can use system (11) with 4 units $\hat{1}$, $\sigma^{kl}$, but with
$\hat{\iota}$-complex coefficients  $u$ and $w_{kl}$:
\begin{equation}
d=u\hat{1}+w_{kl}\sigma^{kl},\hspace{7mm} k,l=1,2,3.
\end{equation}
This subsystem is isomorphic to the biquaternions.
In the same way, multiplication by $\hat{\iota}$ converts $\gamma^{\alpha}$ into $\pi^{\alpha}$,  $\hat{\iota}\gamma^{\alpha}=-\pi^{\alpha}$. 
That is why there is no necessity to use 16 basic units in Dirac's system of numbers (4).  It is enough to use 8 units $\hat{1},
\gamma^\alpha, \sigma^{23}, \sigma^{31}, \sigma^{12}$, but with $\hat{\iota}$-complex coefficients:
\begin{widetext}
\begin{center}{Table 2}\\
\begin{tabular}{|c|c|c|c|}
\hline
&$b\hat{1}$&$b_\mu\gamma^\mu$&$b_{mn}\sigma^{mn}$
\\ \hline
$a\hat{1}$&$ab\hat{1}$&\parbox[c]{3cm}{\begin{center}$a b_\mu\gamma^\mu$\end{center}}
&$ab_{mn}\sigma^{mn}$
\\ \hline
$a_\alpha\gamma^\alpha$&$\breve{b}a_\alpha\gamma^\alpha$&\parbox[c]{5cm}{\begin{center}$a_\alpha \breve{b}^\alpha\hat{1}+\frac{1}{2}([a_k\breve{b}_l]-\hat{\iota}[a_k\breve{b}_l]^\diamond)\sigma^{kl}$\end{center}}
&\parbox[c]{4cm}{\begin{center}$2a^k\breve{b}_{kl}\gamma^l+\hat{\iota}\varepsilon_{kl\alpha\beta}\breve{b}^{kl}a^\alpha\gamma^\beta$\end{center}}
\\ \hline
$a_{kl}\sigma^{kl}$&$ba_{kl}\sigma^{kl}$
&\parbox[c]{4cm}{\begin{center}$-2b^ka_{kl}\gamma^l+\hat{\iota}\varepsilon_{kl\alpha\beta}a^{kl}b^\alpha\gamma^\beta
$\end{center}}
&\parbox[c]{4cm}{\begin{center}$-2a_{kl}b^{kl}\hat{1}+2[a_{km}b_{.l}^{m.}]\sigma^{kl}$
\end{center}}
\\ \hline
\end{tabular}
\end{center}
\end{widetext}
\begin{equation}
D=u\hat{1}+z_\alpha\gamma^\alpha+w_{kl}\sigma^{kl},\hspace{7mm} k,l=1,2,3.
\end{equation}
Here $u, z_\alpha, w_{kl}$ --- $\hat{\iota}$-complex coefficients , $u=u'\hat{1}+u''\hat{\iota}$ etc.  Coefficients $u, z_\alpha, w_{kl}$
are connected with $a, b, c_\alpha, d_\alpha, f_{\alpha\beta}$ in (4) as follows:
\begin{gather}
u=a+\hat{\iota}b,\\
z_\alpha= c_\alpha-\hat{\iota}d_\alpha,\\
w_{kl}=f_{kl}-\hat{\iota}\frac{1}{2}\varepsilon_{kl\alpha\beta}f^{\alpha\beta}=f_{kl}-\hat{\iota}f^\diamond_{kl}.
\end{gather}
Table 2 can be useful for multiplication numbers (16) based on the 8 units with $\hat{\iota}$-complex coefficients.  In Table 2 and in the text below
\begin{equation}
[a_k\breve{b}_l]^\diamond=\frac{1}{2}\varepsilon_{kl\alpha\beta}[a^\alpha\breve{b}^\beta].
\end{equation}
Below, it will be convenient to use the analogy from electrodynamics where electric field $\bm {E}$ and magnetic 
field $\bm{B}$ are used instead of the electromagnetic field tensor $F_{\alpha\beta}$.  
Namely,  instead of 4D vectors or 4D tensors, we can use the parts of 4D vectors $z_\alpha\gamma^\alpha$ or 4D tensors $w_{kl}\sigma^{kl}$  
which behave as scalars and 3D vectors under the action of space rotation.    
 In order to split off these parts let us write  $z_\alpha\gamma^\alpha$ and $w_{kl}\sigma^{kl}$ in the form

\begin{gather}
z_\alpha\gamma^\alpha=z_0\gamma^0+z_k\gamma^k=z^0\gamma^0-z^k\gamma^k\equiv z_0\gamma^0-\bm{z}\bm{\gamma},\\
w_{kl}\sigma^{kl}=2(w_{23}\sigma^{23}+w_{31}\sigma^{31}+w_{12}\sigma^{12})=\nonumber\\
=2(w^{23}\sigma^{23}+w^{31}\sigma^{31}+w^{12}\sigma^{12})\equiv 2\bm{w}\bm{\varsigma}
\end{gather}
 In such 3D $\hat{\iota}$-complex notation, number (16) has a form
\begin{equation}
D=u\hat{1}+z_0\gamma^0-\bm{z}\bm{\gamma}+2\bm{w}\bm{\varsigma}.
\end{equation}
As we will see later, under space rotation, quantities  $u\hat{1}$ and $z_0\gamma^0$ are transformed as $\hat{\iota}$-complex scalars, and quantities
$\bm{z}\bm{\gamma}$ and $\bm{w}\bm{\varsigma}$ are transformed as 3D vectors with $\hat{\iota}$-complex components.  Multiplication of
 $z_0\gamma^0$, $\bm{z}\bm{\gamma}$ and $\bm{w}\bm{\varsigma}$ is reduced  to the 3D dot product and cross product.
 But we need to remember that $\hat{\iota}$ commute with $\bm{\varsigma}$ and anti-commute with  $\bm{\gamma}$. Multiplication
 rules are cited in Table 3.
\begin{gather*}
\mbox{Table 3}\\
\begin{tabular}{|c|c|c|c|c|c|}
\hline
&$b\hat{1}$ &$b\gamma^0$&$\bm{b}\bm{\gamma}$&$\bm{b}\bm{\varsigma}$\\
\hline
$a\hat{1}$& $ab\hat{1}$&\parbox[c]{0.9cm}{\begin{center}$ab\gamma^0$\end{center}}
&\parbox[c]{1.9cm}{\begin{center}$a\bm{b}\bm{\gamma}$\end{center}}
&\parbox[c]{2cm}{\begin{center}$a\bm{b}\bm{\varsigma}$\end{center}}
\\ \hline
$a\gamma^0$& $a\check{b}\gamma^0$&\parbox[c]{0.9cm}{\begin{center}$a\check{b}$\end{center}}
&\parbox[c]{1.9cm}{\begin{center}$-\hat{\iota}a\breve{\bm{b}}\bm{\varsigma}$\end{center}}
&\parbox[c]{2cm}{\begin{center}$-\hat{\iota}a\breve{\bm{b}}\bm{\gamma}$\end{center}}
\\ \hline
$\bm{a}\bm{\gamma}$& $\check{b}\bm{a}\bm{\gamma}$&\parbox[c]{0.9cm}{\begin{center}$\hat{\iota}\breve{b}\bm{a}\bm{\varsigma}$\end{center}}
&\parbox[c]{1.9cm}{\begin{center}$-(\bm{a}\cdot\breve{\bm{b}})+$\\$+(\bm{a}\times\breve{\bm{b}})\bm{\varsigma}$\end{center}}
&\parbox[c]{2cm}{\begin{center}$-\hat{\iota}(\bm{a}\cdot\breve{\bm{b}})\gamma^0+$\\$+(\bm{a}\times\breve{\bm{b}})\bm{\gamma}$\end{center}}
\\ \hline
$\bm{a}\bm{\varsigma}$& $b\bm{a}\bm{\varsigma}$&\parbox[c]{1cm}{\begin{center}$-\hat{\iota}{b}\bm{a}\bm{\gamma}$\end{center}}
&\parbox[c]{2.4cm}{\begin{center}$-\hat{\iota}(\bm{a}\cdot{\bm{b}})\gamma^0+$\\$+(\bm{a}\times{\bm{b}})\bm{\gamma}$\end{center}}
&\parbox[c]{2cm}{\begin{center}$-(\bm{a}\cdot{\bm{b}})+$\\$+(\bm{a}\times{\bm{b}})\bm{\varsigma}$\end{center}}\\ \hline
\end{tabular}
\end{gather*}
\section{Hypercomplex representation of the Lorentz's group}
We need to recall  the rotation description in 3D Euclidian space with the help of quaternions.
This description is possible because unit quaternion $\bm{i}$, $\bm{j}$, $\bm{k}$ commutators coincide with the infinitesimal rotation group
commutators.
In order to describe vector $\bm{a}$ rotation via quaternions we can present vector $\bm{a}$ as a vector-quaternion  \\
\begin{equation}
\bm{a}\bm{q}\equiv a_1\bm{i}+a_2\bm{j}+a_3\bm{k},
\end{equation}
and present rotation by multiplication on the quaternion exponent: $\bm{a}'\bm{q}=e^{\bm{r}\bm{q}}\bm{a}\bm{q}e^{-\bm{r}\bm{q}}$.
Quaternion exponents are determined by their expansion:
\begin{gather}
e^{\pm\bm{r}\bm{q}}=1\pm\bm{r}\bm{q}+\frac{1}{2!}\bm{r}\bm{q}\bm{r}\bm{q}\pm\frac{1}{3!}\bm{r}\bm{q}\bm{r}\bm{q}\bm{r}\bm{q}+\ldots=\nonumber\\
=1\pm\frac{\bm{r}\bm{q}}{r}r-\frac{1}{2!}r^2\mp\frac{1}{3!}\frac{\bm{r}\bm{q}}{r}r^3+\ldots=\cos r\pm\frac{\bm{r}\bm{q}}{r}\sin r.
\end{gather}
Here the vector-quaternion multiplication rule is used
\begin{equation}
\bm{a}\bm{q}\bm{b}\bm{q}=-\bm{a}\cdot\bm{b}+(\bm{a}\times\bm{b})\bm{q}.
\end{equation}
Operation $e^{\bm{r}\bm{q}}\bm{a}\bm{q}e^{-\bm{r}\bm{q}}$ leads to
\begin{gather}
\bm{a}'\bm{q}=(\cos r +\frac{\bm{r}\bm{q}}{r}\sin r)\bm{a}\bm{q}(\cos r-\frac{\bm{r}\bm{q}}{r}\sin r)=\nonumber\\=
\{(\bm{a}\frac{\bm{r}}{r})\frac{\bm{r}}{r}+[\bm{a}\!-\!(\bm{a}\frac{\bm{r}}{r})\frac{\bm{r}}{r}]\cos 2r\!+\!(\frac{\bm{r}\!\times\!\bm{a}}{r})\sin 2r\}\bm{q}.
\end{gather}
As we can see, after the transformation $e^{\bm{r}\bm{q}}\bm{a}\bm{q}e^{-\bm{r}\bm{q}}$, vector  $\bm{a}$ becomes vector $\bm{a}'$:
\begin{gather}
\bm{a}'=(\bm{a}\frac{\bm{r}}{r})\frac{\bm{r}}{r}+[\bm{a}\!-\!(\bm{a}\frac{\bm{r}}{r})\frac{\bm{r}}{r}]\cos 2r+(\frac{\bm{r}}{r}\times\bm{a})\sin 2r.
\end{gather}
The parallel component to the $\bm{r}$  of the vector $\bm{a}$  ($(\bm{a}\frac{\bm{r}}{r})\frac{\bm{r}}{r}$) is not changed, but the perpendicular one
rotates around $\bm{r}$ in the positive direction on the angle $2r$.  The positive direction of the vector  $\bm{r}$ can be determined by the right-hand rule.
Thus, expression $e^{\bm{r}\bm{q}}\bm{a}\bm{q}e^{-\bm{r}\bm{q}}$ describes the rotation of the $\bm{a}$ vector.
Vector $\bm{r}$ is a parameter of this transformation. This rotation is around the $\bm{r}$ vector in the positive direction.  Angle magnitude is equal
to $2 \bm{r}$.

The convenience of the quaternion rotation description is that it gives us a simpler rule of the two rotation  
composition.  Let us write exponent
 $e^{\bm{r}\bm{q}}$ in the following way
\begin{gather}
e^{\bm{r}\bm{q}}=\cos r(1+\frac{\bm{r}\bm{q}}{r}\tan r)=\nonumber\\=
\cos r(1+\bm{\rho}\bm{q})=
\frac{1+\bm{\rho}\bm{q}}{\sqrt{1+\bm{\rho}\cdot\bm{\rho}}}.
\end{gather}
Here
\begin{gather}
\bm{\rho}=\frac{\bm{r}}{r}\tan r.
\end{gather}
By replacing $\bm{r}$ with $\bm{\rho}$ we change the rotation parametrization.  Vector  $\bm{\rho}$  becomes the rotation parameter.
Vector $\bm{a}$ rotates around  $\bm{\rho}$  in the positive direction and the length of $\bm{\rho}$ is equal to the tangent of the half-angle rotation:
\begin{gather}
\tan r=\sqrt{\bm{\rho}\cdot\bm{\rho}}.
\end{gather}
As we will see the composition rule becomes very simple if parameter $\bm{\rho}$ is used. 
Let us describe the first rotation by exponent $e^{\bm{r}_1\bm{q}}$  and the following rotation by exponent $e^{\bm{r}_2\bm{q}}$.  Then their 
composition can be written as multiplication of the two exponents $e^{\bm{r}_1\bm{q}}$  and  $e^{\bm{r}_2\bm{q}}$:  
\begin{gather}
e^{\bm{r}_2\bm{q}}e^{\bm{r}_1\bm{q}}=
\frac{(1+\bm{\rho}_2\bm{q})(1+\bm{\rho}_1\bm{q})}{\sqrt{1+\bm{\rho}_2\cdot\bm{\rho}_2}\sqrt{1+\bm{\rho}_1\cdot\bm{\rho}_1}}=\nonumber\\=
\frac{1-\bm{\rho}_1\cdot\bm{\rho}_2}{\sqrt{(1+\bm{\rho}_2\cdot\bm{\rho}_2)(1+\bm{\rho}_1\cdot\bm{\rho}_1)}}
\Bigl(1+\frac{\bm{\rho}_1+\bm{\rho}_2+\bm{\rho}_2\times\bm{\rho}_1}{1-\bm{\rho}_1\cdot\bm{\rho}_2}\bm{q}\Bigr).
\end{gather}
Let us set
\begin{gather}
\bm{\rho}=\frac{\bm{\rho}_1+\bm{\rho}_2+\bm{\rho}_2\times\bm{\rho}_1}{1-\bm{\rho}_1\cdot\bm{\rho}_2}.
\end{gather}
It is easy to see that
\begin{gather}
\bm{\rho}\cdot\bm{\rho}=
\frac{(1+\bm{\rho}_2\cdot\bm{\rho}_2)(1+\bm{\rho}_1\cdot\bm{\rho}_1)}{(1-\bm{\rho}_1\cdot\bm{\rho}_2)^2}-1.
\end{gather}
That is why
\begin{gather}
e^{\bm{r}_2\bm{q}}e^{\bm{r}_1\bm{q}}=
\frac{1+\bm{\rho}\bm{q}}{\sqrt{1+\bm{\rho}\cdot\bm{\rho}}}=e^{\bm{r}\bm{q}}.
\end{gather}
Here
\begin{gather}
\bm{r}=\bm{\rho}\frac{r}{\tan r},\hspace{7mm}\tan r=\sqrt{\bm{\rho}\cdot\bm{\rho}}.
\end{gather}
Thus, parameter $\bm{\rho}$ of the general transformation $e^{\bm{r}\bm{q}}$, which is the result of rotation $e^{\bm{r}_1\bm{q}}$
and further rotation of $e^{\bm{r}_2\bm{q}}$, is determined by the composition rule (33).  We can see that this rule differs from the rule 
(1) by the transformation parameter only. Parameter  $\bm{\rho}$ is the real number while parameter $\bm{c}$ is the complex number.

Let us proceed to Loretz's transformation.

Hypercomplex units $\sigma^{\alpha\beta}$ commutators coincide with the infinitesimal Lorentz's group operator commutators.
That is why in the hypercomplex description proper Lorentz's transformations can be presented as
\begin{equation}
D'=e^{\frac{1}{2}L_{\alpha\beta}\sigma^{\alpha\beta}}De^{-\frac{1}{2}L_{\alpha\beta}\sigma^{\alpha\beta}}.
\end{equation}
Exponent $e^{L_{\alpha\beta}\sigma^{\alpha\beta}}$ is determined by the expansion
\begin{gather}
e^{\frac{1}{2}L_{\alpha\beta}\sigma^{\alpha\beta}}=\hat{1}+\frac{1}{2}L_{\alpha\beta}\sigma^{\alpha\beta}+
\frac{1}{2!}\frac{1}{2}L_{\alpha\beta}\sigma^{\alpha\beta}\frac{1}{2}L_{\alpha\beta}\sigma^{\alpha\beta}+\ldots
\end{gather}
Product $(L_{\alpha\beta}\sigma^{\alpha\beta}L_{\alpha\beta}\sigma^{\alpha\beta})/4$ in (38) is a $\hat{\iota}$-complex number:
\begin{gather}
\frac{1}{2}L_{\alpha\beta}\sigma^{\alpha\beta}\frac{1}{2}L_{\alpha\beta}\sigma^{\alpha\beta}=\nonumber\\=
-\frac{1}{2}(L^{\alpha\beta}L_{\alpha\beta}-\hat{\iota}L^{\alpha\beta}L^\diamond_{\alpha\beta})=
-\frac{1}{2}L^{\alpha\beta}\!\!\bullet L_{\alpha\beta}\equiv L^2.
\end{gather}
Correspondingly for the exponent  (38) we get
\begin{gather}
e^{\frac{1}{2}L_{\alpha\beta}\sigma^{\alpha\beta}}=\cosh L+\frac{\sinh L}{L}\frac{1}{2}L_{\alpha\beta}\sigma^{\alpha\beta}=\nonumber\\
=\cosh L\bigl(1+\frac{\tanh L}{L}\frac{1}{2}L_{\alpha\beta}\sigma^{\alpha\beta}\bigr)
=\frac{1+\frac{1}{2}\Lambda_{\alpha\beta}\sigma^{\alpha\beta}}
{\sqrt{1+\frac{1}{2}\Lambda^{\alpha\beta}\!\!\bullet \Lambda_{\alpha\beta}}}.
\end{gather}
Here $\Lambda_{\alpha\beta}=\frac{\tanh L}{L}L_{\alpha\beta}$, $L=\sqrt{L^2}$, $L^2$ and $L$ are $\hat{\iota}$-complex numbers.
We can use either root sign in the  definition of $L$ because $\sinh L/L$ is an even function.   Obviously,
\begin{gather}
e^{\frac{1}{2}L_{\alpha\beta}\sigma^{\alpha\beta}}e^{-\frac{1}{2}L_{\alpha\beta}\sigma^{\alpha\beta}}=
\bigl(\cosh L+\frac{1}{2}\frac{L_{\alpha\beta}\sigma^{\alpha\beta}}{L}\sinh L\bigr)\times\nonumber\\\times
\bigl(\cosh L-\frac{1}{2}\frac{L_{\alpha\beta}\sigma^{\alpha\beta}}{L}\sinh L\bigr)
=\cosh^2 L - \sinh^2 L=1.
\end{gather}
With the help of table (1) and exponent (40) it is easy to show that equation (37) expresses the proper Lorentz's transformation if  we 
present scalars and pseudo-scalars as numbers $a$ and $a\hat{\iota}$, vectors and pseudo-vectors as numbers  $a_\alpha\gamma^\alpha$ and $a_\alpha\pi^\alpha$,
and antisymmetric tensors of the second order as numbers $a_{\alpha\beta}\sigma^{\alpha\beta}$.
 In fact, this operation does not change numbers $a\hat{1}$ and $a\hat{\iota}$:
\begin{gather}
a'\hat{1}=e^{\frac{1}{2}L_{\alpha\beta}\sigma^{\alpha\beta}}a\hat{1}e^{-\frac{1}{2}L_{\alpha\beta}\sigma^{\alpha\beta}}\!=\nonumber\\=
e^{\frac{1}{2}L_{\alpha\beta}\sigma^{\alpha\beta}}e^{-\frac{1}{2}L_{\alpha\beta}\sigma^{\alpha\beta}}a\hat{1}\!=a\hat{1},\\
a'\hat{\iota}=e^{\frac{1}{2}L_{\alpha\beta}\sigma^{\alpha\beta}}a\hat{\iota}e^{-\frac{1}{2}L_{\alpha\beta}\sigma^{\alpha\beta}}\!=\nonumber\\=
e^{\frac{1}{2}L_{\alpha\beta}\sigma^{\alpha\beta}}e^{-\frac{1}{2}L_{\alpha\beta}\sigma^{\alpha\beta}}a\hat{\iota}=a\hat{\iota},
\end{gather}
and does not change quadratic forms $a^\alpha a_\alpha=a^2_0-a^2_1-a^2_2-a^2_3$, 
which are connected to vectors and pseudo-vectors:
\begin{gather}
a'^\alpha a'_\alpha=a'_\alpha\gamma^\alpha a'_\alpha\gamma^\alpha=\nonumber\\
=e^{\frac{1}{2}L_{\alpha\beta}\sigma^{\alpha\beta}}a_\alpha\gamma^\alpha e^{-\frac{1}{2}L_{\alpha\beta}\sigma^{\alpha\beta}}
e^{\frac{1}{2}L_{\alpha\beta}\sigma^{\alpha\beta}}a_\alpha\gamma^\alpha e^{-\frac{1}{2}L_{\alpha\beta}\sigma^{\alpha\beta}}=\nonumber\\
=e^{\frac{1}{2}L_{\alpha\beta}\sigma^{\alpha\beta}}a_\alpha\gamma^\alpha a_\alpha\gamma^\alpha e^{-\frac{1}{2}L_{\alpha\beta}\sigma^{\alpha\beta}}=
a^\alpha a_\alpha,\\
a'^\alpha a'_\alpha=a'_\alpha\pi a'_\alpha\pi=\nonumber\\
=e^{\frac{1}{2}L_{\alpha\beta}\sigma^{\alpha\beta}}a_\alpha\pi e^{-\frac{1}{2}L_{\alpha\beta}\sigma^{\alpha\beta}}
e^{\frac{1}{2}L_{\alpha\beta}\sigma^{\alpha\beta}}a_\alpha\pi e^{-\frac{1}{2}L_{\alpha\beta}\sigma^{\alpha\beta}}=\nonumber\\
=e^{\frac{1}{2}L_{\alpha\beta}\sigma^{\alpha\beta}}a_\alpha\pi^\alpha a_\alpha\pi^\alpha e^{-\frac{1}{2}L_{\alpha\beta}\sigma^{\alpha\beta}}=
a^\alpha a_\alpha,
\end{gather}
Also it does not change quadratic forms $a^{\alpha\beta}a_{\alpha\beta}$ and $a^{\alpha\beta}a^\diamond_{\alpha\beta}$, which are connected to 
antisymmetric tensors:
\begin{gather}
2a'^{\alpha\beta}\!\!\bullet a'_{\alpha\beta}\!=\!2(a'^{\alpha\beta}a'_{\alpha\beta}\!-\hat{\iota}a'^{\alpha\beta}a'^\diamond_{\alpha\beta})
=\!-a'_{\alpha\beta}\sigma^{\alpha\beta}a'_{\alpha\beta}\sigma^{\alpha\beta}\!=\nonumber\\
=\!-e^{\frac{1}{2}L_{\alpha\beta}\sigma^{\alpha\beta}}a_{\alpha\beta}\sigma^{\alpha\beta}e^{-\frac{1}{2}L_{\alpha\beta}\sigma^{\alpha\beta}}\times
\nonumber\\
\times e^{\frac{1}{2}L_{\alpha\beta}\sigma^{\alpha\beta}}\!a_{\alpha\beta}\sigma^{\alpha\beta}e^{-\frac{1}{2}L_{\alpha\beta}\sigma^{\alpha\beta}}\!=\nonumber\\
=-e^{\frac{1}{2}L_{\alpha\beta}\sigma^{\alpha\beta}}a_{\alpha\beta}\sigma^{\alpha\beta}a_{\alpha\beta}\sigma^{\alpha\beta}
e^{-\frac{1}{2}L_{\alpha\beta}\sigma^{\alpha\beta}}=\nonumber\\
=2(a^{\alpha\beta}a_{\alpha\beta}-\hat{\iota}a^{\alpha\beta}a^\diamond_{\alpha\beta})=2a^{\alpha\beta}\!\!\bullet a_{\alpha\beta}.
\end{gather}
Quadratic forms $a^{\alpha\beta}a_{\alpha\beta}$ and $a^{\alpha\beta}a^\diamond_{\alpha\beta}$ are counterparts of the 
 scalar and pseudo-scalar invariants  $\bm{E}^2-c^2\bm{B}^2$ and $\bm{E}\cdot\bm{B}$ well-known from electrodynamics.

The conservation law of  $a_\alpha a^\alpha$ proves that transformation (37) is the Lorentz's transformation of the vectors and pseudo-vectors. 
In turn, the conservation law of  $a^{\alpha\beta}a_{\alpha\beta}$ and $a^{\alpha\beta}a^\diamond_{\alpha\beta}$ proves
that transformation (37) is Lorentz's transformation of the antisymmetric tensors.  Six components of the antisymmetric tensor  of the second order  
$L_{\alpha\beta}\sigma^{\alpha\beta}$ are parameters of this transformation.

Instead of $L_{\alpha\beta}$  we can use the dimensionless variable  $L_{\alpha\beta}=\frac{q}{mc}F_{\alpha\beta}\delta t$,
which is proportional to the electromagnetic field tensor  $F_{\alpha\beta}$, where $\delta\tau$ is an infinitesimal proper time interval of a particle of mass  $m$ and charge
  $q$.  In this case with linear accuracy on $\delta\tau$ terms for the 4-impulse transformations of the particle we get 
\begin{gather}
p'_\alpha\gamma^\alpha
=e^{\frac{1}{2}L_{\alpha\beta}\sigma^{\alpha\beta}}p_\alpha\gamma^\alpha e^{-\frac{1}{2}L_{\alpha\beta}\sigma^{\alpha\beta}}=\nonumber\\
=(\hat{1}+\frac{1}{2}L_{\alpha\beta}\sigma^{\alpha\beta})p_\alpha\gamma^\alpha(\hat{1}-\frac{1}{2}L_{\alpha\beta}\sigma^{\alpha\beta})=\nonumber\\
=(p_\alpha+L_{\alpha\beta}p^\beta)\gamma^\alpha=(p_\alpha+\frac{q}{mc}F_{\alpha\beta}p^\beta\delta t)\gamma^\alpha=\nonumber\\
=(p_\alpha+\delta p_\alpha)\gamma^\alpha.
\end{gather}
The change in momentum in this case is the same as its change under the action of the Lorentz force:
\begin{gather}
\frac{dp_\alpha}{d\tau}=\frac{q}{mc}F_{\alpha\beta}p^\beta.
\end{gather}
In other words, the moment of the charged particle in the electromagnetic field experiences the sequence of
infinitesimal Lorentz transformations.  Parameters of these transformations  ($L_{\alpha\beta}$) are proportional to the the
electromagnetic filed tensor $F_{\alpha\beta}$. We can also treat motion of any particle with constant mass as a sequence of
infinitesimal Lorentz transformations, because in this case, quantity $p_\alpha p^\alpha$ is conserved.


Let us return to the main topic of the investigation.  
It is also clear that $e^{\frac{1}{2}L_{\alpha\beta}\sigma^{\alpha\beta}}\psi$ is Lorentz's transformation of the Dirac bispinors $\psi$,
because it does not change bilinear combination  $\bar{\psi}\psi$. In fact, because of
\begin{gather}
(L_{\alpha\beta}\sigma^{\alpha\beta})^\dag\gamma^0=-\gamma^0 L_{\alpha\beta}\sigma^{\alpha\beta},
\end{gather}
we get
\begin{gather}
\bar{\psi'}=\psi'^\dag\gamma^0=(e^{\frac{1}{2}L_{\alpha\beta}\sigma^{\alpha\beta}}\psi)^\dag\gamma^0=
\psi^\dag(e^{\frac{1}{2}L_{\alpha\beta}\sigma^{\alpha\beta}})^\dag\gamma^0=\nonumber\\=
\psi^\dag\gamma^0e^{-\frac{1}{2}L_{\alpha\beta}\sigma^{\alpha\beta}}=\bar{\psi}e^{-\frac{1}{2}L_{\alpha\beta}\sigma^{\alpha\beta}}.
\end{gather}
That is why
\begin{gather}
\bar{\psi'}\psi'=\bar{\psi}e^{-\frac{1}{2}L_{\alpha\beta}\sigma^{\alpha\beta}}e^{\frac{1}{2}L_{\alpha\beta}\sigma^{\alpha\beta}}\psi=
\bar{\psi}\psi.
\end{gather}
We have seen, therefore, that expression (37) performs an arbitrary proper Lorentz's transformation of the vectors and antisymmetric tensors of the
second order.  The product  $e^{\frac{1}{2}L_{\alpha\beta}\sigma^{\alpha\beta}}\psi$ performs an arbitrary proper Lorentz's transformation of the
bispinors.  Transformation of all these quantities is performed by the same exponents $e^{\pm\frac{1}{2}L_{\alpha\beta}\sigma^{\alpha\beta}}$.
This considerably simplifies treatment in many cases.

Instead of hypercomplex numbers (4) with 16 basic units we can use  numbers (16) with 8 units and $\hat{\iota}$-complex coefficients.  In this
case, the expression for the exponent (38) and subsequent expressions becomes simpler, because 
instead of tensor  $L_{\alpha\beta}\sigma^{\alpha\beta}$
with 6 basic units  $\sigma^{\alpha\beta}$ and real $L_{\alpha\beta}$ we must use tensor  $l_{kl}\sigma^{kl}$ with 3 basic units $\sigma^{kl}$ and $\hat{\iota}$-complex $l_{kl}$.  Proper Lorentz's transformations become:
\begin{equation}
D'=e^{\frac{1}{2}l_{kl}\sigma^{kl}}De^{-\frac{1}{2}l_{kl}\sigma^{kl}},
\end{equation}
where $\frac{1}{2}l _{kl}\sigma^{kl}=l_{23}\sigma^{23}+l_{31}\sigma^{31}+l_{12}\sigma^{12}$.
Exponent $e^{\frac{1}{2}l_{kl}\sigma^{kl}}$ equals
\begin{gather}
e^{\frac{1}{2}l_{kl}\sigma^{kl}}=\cosh l+\frac{\sinh l}{l}\frac{1}{2}l_{kl}\sigma^{kl}=\nonumber\\
=\cosh l\bigl(1+\frac{\tanh l}{l}\frac{1}{2}l_{kl}\sigma^{kl}\bigr)
=\frac{1+\frac{1}{2}\lambda_{kl}\sigma^{kl}}
{\sqrt{1+\frac{1}{2}\lambda^{kl}\lambda_{kl}}}.
\end{gather}
Here $l$ --- $\hat{\iota}$-complex number
\begin{gather}
l=\sqrt{\frac{1}{2}l_{kl}\sigma^{kl}\frac{1}{2}l_{kl}\sigma^{kl}}=\sqrt{-\frac{1}{2}l^{kl}l_{kl}},
\end{gather}
and $\lambda_{kl}=l_{kl}\frac{\tanh l}{l}$.  The sign in front of  $l$ is not important because $\sinh l/l$ is an even function.
Expression (53) is simpler than (40) because regular tensor contraction  $l^{kl}l_{kl}$ of $\hat{\iota}$-complex parameters $l_{kl}$ is used
instead of (39), which determines $\hat{\iota}$-complex number $L^{\alpha\beta}\bullet L_{\alpha\beta}$.

Vectors and pseudo-vectors are represented by $z_\alpha\gamma^\alpha$ numbers (with $\hat{\iota}$-complex $z_\alpha$) and
tensors are represented by $w_{kl}\sigma^{kl}$ (with $\hat{\iota}$-complex $w_\alpha$) when expression (52) is used for
proper Lorentz's transformations.

We can achieve further simplifications for the exponent (38) and subsequent expressions if we use 
 3D $\hat{\iota}$-complex notation (23).   In this case, the exponent power in (38) is
\begin{gather}
\frac{L_{\alpha\beta}\sigma^{\alpha\beta}}{2}=\frac{l_{kl}\sigma^{kl}}{2}=
l^{23}\sigma^{23}+l^{31}\sigma^{31}+l^{12}\sigma^{12}\equiv\bm{l}\bm{\varsigma},
\end{gather}
Proper Lorentz's transformation (37) is presented as the product:
\begin{gather}
D'=e^{\bm{l}\bm{\varsigma}}De^{-\bm{l}\bm{\varsigma}}
\end{gather}
with $e^{\pm\bm{l}\bm{\varsigma}}$ exponent
\begin{gather}
e^{\pm\bm{l}\bm{\varsigma}}=\hat{1}\pm\bm{l}\bm{\varsigma}+\frac{1}{2!}(\bm{l}\bm{\varsigma}\bm{l}\bm{\varsigma})\pm\ldots=
\cosh l\pm\frac{\sinh l}{l}\bm{l}\bm{\varsigma}=\nonumber\\=
\cosh l(1\pm\frac{\tanh l}{l}\bm{l}\bm{\varsigma})=\frac{1\pm\bm{\lambda}\bm{\varsigma}}{\sqrt{1+\bm{\lambda}\cdot\bm{\lambda}}}.
\end{gather}
Here $\bm{l}$ and $\bm{\lambda}$ are $\hat{\iota}$-complex 3D-vectors, and $l$ is an $\hat{\iota}$-complex scalar:
\begin{gather}
\bm{l}=\bm{r}+\hat{\iota}\bm{b},\\
l=\sqrt{\bm{l}\bm{\varsigma}\bm{l}\bm{\varsigma}}=
\sqrt{-\bm{l}\cdot\bm{l}}=\sqrt{\bm{b}\cdot\bm{b}-\bm{r}\cdot\bm{r}-2\hat{\iota}\bm{r}\cdot\bm{b}},\\
\bm{\lambda}=\frac{\tanh l}{l}\bm{l}.
\end{gather}
Use of the expression (23) assumes that 4D vectors and 4D pseudo-vectors are represented by numbers $a_0\gamma^0-\bm{a}\bm{\gamma}$
(with $\hat{\iota}$-complex coefficients)  and antisymmetric tensors of the second order are represented by $\bm{a}\bm{\varsigma}$ numbers
(with $\hat{\iota}$-complex coefficients).

Use of these expressions gives us definite formulae for arbitrary Lorentz's transformations, in particular,  
rotations and boosts.

\subsection{Transformation of vectors and pseudo-vectors}
Proper Lorentz's transformations change vectors and pseudo-vectors in the same way.  That is why we can treat them jointly.
Let us discuss the change of the hypercomplex number
\begin{gather}
z_\alpha\gamma^\alpha=z_0\gamma^0+z_1\gamma^1+z_2\gamma^2+z_3\gamma^3=\nonumber\\
=z_0\gamma^0-z^1\gamma^1-z^2\gamma^2-z^3\gamma^3=z_0\gamma^0-\bm{z}\bm{\gamma}
\end{gather}
by  (56). Here $z_\alpha$ are  $\hat{\iota}$-complex components of the $\hat{\iota}$-complex 4D vector $z_\alpha$.
Numbers $z_\alpha\gamma^\alpha$ are vectors if $z_\alpha$ are real and $z_\alpha\gamma^\alpha$ are pseudo-vectors if $z_\alpha$
are $\hat{\iota}$-complex.
\begin{gather}
z'_\alpha\gamma^\alpha=z'_0\gamma^0-\bm{z}'\bm{\gamma}=
e^{\bm{l}\bm{\varsigma}}z_\alpha\gamma^\alpha e^{-\bm{l}\bm{\varsigma}}=\nonumber\\=
(\cosh{l}+\frac{\bm{l}}{l}\bm{\varsigma}\sinh{l})(z_0\gamma^0-\bm{z}\bm{\gamma})
(\cosh{l}-\frac{\bm{l}}{l}\bm{\varsigma}\sinh{l})=\nonumber\\=
\Bigl\{z_0\Bigl[\cosh{l}\cosh\breve{\bm{l}}+\bigl(\frac{\bm{l}}{l}\cdot\frac{\breve{\bm{l}}}{\breve{l}}\bigr)\sinh{l}\sinh\breve{l}\Bigr]+\nonumber\\
+\bm{z}\cdot\Bigl[\hat{\iota}(\frac{\bm{l}}{l}\times\frac{\breve{\bm{l}}}{\breve{l}})\sinh{l}\sinh\breve{l}+\nonumber\\
+\hat{\iota}\Bigl(\frac{\bm{l}}{l}\sinh{l}\cosh\breve{l}-
\frac{\breve{\bm{l}}}{\breve{l}}\sinh\breve{l}\cosh l\Bigr)\Bigr]\Bigr\}\gamma^0-\nonumber\\-
\Bigl\{\bm{z}\cosh\bm{l}\cosh\breve{l}
-\bm{z}\!\times\!\Bigl(\frac{\bm{l}}{l}\sinh l\cosh\breve{l}+\frac{\breve{\bm{l}}}{\breve{l}}\sinh\breve{l}\cosh l\Bigr)+\nonumber\\
+\Bigl[\frac{\bm{l}}{l}\bigl(\bm{z}\cdot\frac{\breve{\bm{l}}}{\breve{l}}\bigr)+\frac{\breve{\bm{l}}}{\breve{l}}\bigl(\bm{z}\cdot\frac{\bm{l}}{l}\bigr)
-\bm{z}\bigl(\frac{\bm{l}}{l}\cdot\frac{\breve{\bm{l}}}{\breve{l}}\bigr)\Bigr]\sinh{l}\sinh\breve{l}+\nonumber\\+
z_0\hat{\iota}\Bigl(\frac{\bm{l}}{l}\sinh l\cosh\breve{l}-\frac{\breve{\bm{l}}}{\breve{l}}\sinh\breve{l}\cosh{l}\Bigr)-\nonumber\\
-z_0\hat{\iota}\bigl(\frac{\bm{l}}{l}\times\frac{\breve{\bm{l}}}{\breve{l}}\bigr)\sinh{l}\sinh\breve{l}\Bigr\}\bm{\gamma}.
\end{gather}
The coefficients in front of $z_\alpha$ are real. That is why the transformations of the real and complex parts of the $\hat{\iota}$-complex vector $z_\alpha$
are independent and similar.  Just a reminder that vector $z_\alpha$ describes vectors and pseudo-vectors.

In general case, expression (62) (when $\bm{l}$ is an $\hat{\iota}$-complex value) describes arbitrary proper Lorentz's
transformations of vectors and pseudo-vectors. In particular, it can be  rotations and boosts.
\subsection{Rotations}
When $\bm{b}=0$ and $\bm{l}=\bm{r}$, then
$l=\sqrt{\bm{r}\bm{\varsigma}_2\bm{r}\bm{\varsigma}_2}=\sqrt{-\bm{r}\cdot\bm{r}}=\hat{\iota}\sqrt{\bm{r}\cdot\bm{r}}=\hat{\iota}r$,
$\sinh l=\hat{\iota}\sin r$, $\cosh l=\cos r$, matrix $e^{\bm{r}\bm{\varsigma}_2}$ is unitary:
$(e^{\bm{r}\bm{\varsigma}})^\dag=e^{-\bm{r}\bm{\varsigma}}=(e^{\bm{r}\bm{\varsigma}})^{-1}$.
So, transformation (62) becomes
\begin{gather}
z'_0\gamma^0-\bm{z}'\bm{\gamma}=e^{\bm{r}\bm{\varsigma}}z_\alpha\gamma^\alpha e^{-\bm{r}\bm{\varsigma}}=\nonumber\\=
(\cos r\!+\!\frac{\bm{r}\bm{\varsigma}}{r}\sin r)(z_0\gamma^0\!-\!\bm{z}\bm{\gamma})(\cos r\!-\!\frac{\bm{r}\bm{\varsigma}}{r}\sin r)=
z_0\gamma^0-\nonumber\\-\Bigl[\frac{\bm{r}}{r}(\frac{\bm{r}\cdot\bm{z}}{r})
+\bigl[\bm{z}-\frac{\bm{r}}{r}(\frac{\bm{r}\cdot\bm{z}}{r})\bigr]\cos2r+(\frac{\bm{r}\times\bm{z}}{r})\sin 2r\Bigr]\bm{\gamma}.
\end{gather}
I.e.
\begin{gather}
z'_0=z_0,\\
\bm{z}'=\frac{\bm{r}}{r}(\frac{\bm{r}\cdot\bm{z}}{r})
+\bigl[\bm{z}-\frac{\bm{r}}{r}(\frac{\bm{r}\cdot\bm{z}}{r})\bigr]\cos2r+(\frac{\bm{r}\times\bm{z}}{r})\sin 2r.
\end{gather}
As we can see this transformation does not change the time component $z_0$ of the  4D vector  $z^\alpha$ and
rotates its spatial part $\bm{z}$ around the $\bm{r}$ vector in the positive direction on an angle of $2r$.
\subsection{Boosts}
When $\bm{r}=0$ and $\bm{l}=\hat{\iota}\bm{b}$, then $\bm{l}\bm{\varsigma}_2=\hat{\iota}\bm{b}\bm{\varsigma}_2$, $l=\sqrt{\bm{l}\bm{\varsigma}_2\bm{l}\bm{\varsigma}_2}=\sqrt{\bm{b}\cdot\bm{b}}=b$, $\sinh l=\sinh b$, $\cosh l=\cosh b$, matrix
$e^{\hat{\iota}\bm{b}\bm{\varsigma}_2}$ is hermitian:
$(e^{\hat{\iota}\bm{b}\bm{\varsigma}_2})^\dag=e^{\hat{\iota}\bm{b}\bm{\varsigma}_2}$.
So, transformation (62) becomes
\begin{gather}
z'_\alpha\gamma^\alpha=z'_0\gamma^0-\bm{z}'\bm{\gamma}=
e^{\hat{\iota}\bm{b}\bm{\varsigma_2}}z_\alpha\gamma^\alpha e^{-\hat{\iota}\bm{b}\bm{\varsigma_2}}=\nonumber\\=
(\cosh b+\frac{\hat{\iota}\bm{b}\bm{\varsigma}}{b}\sinh b)(z_0\gamma^0-\bm{z}\bm{\gamma})
(\cosh b-\frac{\hat{\iota}\bm{b}\bm{\varsigma}}{b}\sinh b)=\nonumber\\=
\Bigl[z_0\cosh 2b-(\frac{\bm{z}\cdot\bm{b}}{b})\sinh 2b\Bigr]\gamma^0-\nonumber\\-
\Bigl[\bm{z}+(\frac{\bm{z}\cdot\bm{b}}{b})\frac{\bm{b}}{b}(\cosh 2b-1)-z_0\frac{\bm{b}}{b}\sinh 2b\Bigr]\bm{\gamma}.
\end{gather}
Let us set up $\frac{\bm{b}}{b}=\frac{\bm{v}}{v}$, $\tanh2b=\frac{v}{c}$. In this case expression (66) describes the transformation of the  contravariant
components of the vector $z^\alpha$ under the transition from the reference frame  $S$ to the reference frame $S'$. Reference frame $S'$  
is moving with relative velocity $\bm{v}$ with respect to  $S$.
 Also  $\cosh 2b=1/\sqrt{1-\frac{v^2}{c^2}}$, $\sinh 2b=\frac{v}{c}/\sqrt{1-\frac{v^2}{c^2}}$ and instead of (66) we get
\begin{gather}
z'_0\gamma^0-\bm{z}'\bm{\gamma}=
\Bigl[z_0-(\frac{\bm{z}\cdot\bm{v}}{v})\Bigr]\frac{1}{\sqrt{1-\frac{v^2}{c^2}}}\gamma^0-\nonumber\\-
\Bigl[\bm{z}+\frac{(\bm{z}\cdot\bm{v})\bm{v}}{v^2}\Bigl(\frac{1}{\sqrt{1-\frac{v^2}{c^2}}}-1\Bigr)-\frac {z_0}{\sqrt{1-\frac{v^2}{c^2}}}\frac{\bm{v}}{c}\Bigr]\bm{\gamma}.
\end{gather}
Thus
\begin{gather}
z'_0=\Bigl[z_0-\frac{(\bm{z}\cdot\bm{v})}{c}\Bigr]\frac{1}{\sqrt{1-\frac{v^2}{c^2}}},\\
\bm{z}'=\bm{z}+\frac{(\bm{z}\cdot\bm{v})\bm{v}}{v^2}\Bigl(\frac{1}{\sqrt{1-\frac{v^2}{c^2}}}-1\Bigr)-\frac{z_0}{\sqrt{1-\frac{v^2}{c^2}}}\frac{\bm{v}}{c}.
\end{gather}
This is well-known $z^\alpha$ vector transformation caused by the boost, i.e. transformation from the reference frame $S$ to the reference frame $S'$.
\subsection{Tensor transformation}
Using $F_{\alpha\beta}$ numbers we can represent an antisymmetric tensor of the second order as
\begin{gather}
\frac{1}{2}F_{\alpha\beta}\sigma^{\alpha\beta}=\nonumber\\=
E_x\sigma^{01}\!+E_y\sigma^{02}\!+E_z\sigma^{03}\!-c(B_x\sigma^{23}\!+B_y\sigma^{31}\!+B_z\sigma^{12})=\nonumber\\=
-(\hat{\iota}\bm{E}+c\bm{B})\bm{\varsigma}=\bm{F}\bm{\varsigma}.
\end{gather}
Here, the usual designations are used for the electromagnetic field tensor
\begin{gather*}
F^{01},F^{01},F^{01}=-E_x,-E_y,-E_z,\nonumber\\F^{23},F^{31},F^{12}=-cB_x,-cB_y,-cB_z,
\end{gather*}
also, $\hat{\iota}$-complex numbers are introduced 
\begin{gather}
\bm{F}=-(\hat{\iota}\bm{E}+c\bm{B}).
\end{gather}
The Lorentz's transformation of the antisymmetric tensor of the second order are introduced by the same expression (56) that is used for the vector
transformation.
In fact 
\begin{gather}
\frac{1}{2}F'_{\alpha\beta}\sigma^{\alpha\beta}=\bm{F}'\bm{\varsigma}=-(\hat{\iota}\bm{E}'+c\bm{B}')\bm{\varsigma}=
e^{\bm{l}\bm{\varsigma}}\frac{1}{2}F_{\alpha\beta}\sigma^{\alpha\beta}e^{-\bm{l}\bm{\varsigma}}=\nonumber\\=
(\cosh l+\frac{\bm{l}\bm{\varsigma}}{l}\sinh l)\bm{F}\bm{\varsigma}
(\cosh l-\frac{\bm{l}\bm{\varsigma}}{l}\sinh l)=\nonumber\\=
\Bigl\{-\frac{\bm{l}}{l}\Bigl(\frac{\bm{F}\!\cdot\!\bm{l}}{l}\Bigr)\!+\!\Bigl[\bm{F}+
\frac{\bm{l}}{l}\Bigl(\frac{\bm{F}\!\cdot\!\bm{l}}{l}\Bigr)\Bigr]\cosh 2l\!+\!
\frac{\bm{l}\!\times\!\bm{F}}{l}\sinh 2l\Bigr\}\bm{\varsigma}.
\end{gather}
Here the convenient equality is used
\begin{gather}
\bm{c}\bm{\varsigma}\bm{a}\bm{\varsigma}\bm{c}\bm{\varsigma}=[\bm{a}c^2-2\bm{c}(\bm{a}\cdot\bm{c})]\bm{\varsigma}.
\end{gather}
So, expression  $e^{\bm{l}\bm{\varsigma}}\frac{1}{2}F_{\alpha\beta}\sigma^{\alpha\beta}e^{-\bm{l}\bm{\varsigma}}=
e^{\bm{l}\bm{\varsigma}}\bm{F}\bm{\varsigma}e^{-\bm{l}\bm{\varsigma}}$ changes the $\hat{\iota}$-complex number $\bm{F}$ in
the following way:
\begin{gather}
\bm{F}'=
-\frac{\bm{l}}{l}\Bigl(\bm{F}\!\cdot\!\frac{\bm{l}}{l}\Bigr)+\Bigl[\bm{F}+
\frac{\bm{l}}{l}\Bigl(\bm{F}\!\cdot\!\frac{\bm{l}}{l}\Bigr)\Bigr]\cosh 2l+
\frac{\bm{l}}{l}\!\times\!\bm{F}\sinh 2l.
\end{gather}
For the rotations, when $\bm{l}=\bm{r}$, $l=\hat{\iota}r$ we get
\begin{gather}
\bm{F}'=\frac{\bm{r}}{r}\Bigl(\bm{F}\!\cdot\!\frac{\bm{r}}{r}\Bigr)+\Bigl[\bm{F}-
\frac{\bm{r}}{r}\Bigl(\bm{F}\!\cdot\!\frac{\bm{r}}{r}\Bigr)\Bigr]\cos 2r+\frac{\bm{r}}{r}\!\times\!\bm{F}\sin 2r.
\end{gather}
This is a rotation of the vectors $\bm{E}$ and $c\bm{B}$ in the positive direction around the vector $\bm{r}$ on an angle of $2r$.

For the Lorentz boost when $\bm{l}=\hat{\iota}\bm{b}$, $l=b$, $\frac{\bm{b}}{b}=\frac{\bm{v}}{v}$, $\tanh 2b=\frac{v}{c}$, the
$\hat{\iota}$-complex vector $\bm{F}$ is transformed as
\begin{gather}
\bm{F}'=
\frac{\bm{v}}{v}\Bigl(\bm{F}\cdot\frac{\bm{v}}{v}\Bigr)+
\frac{\bm{F}-\frac{\bm{v}}{v}\Bigl(\bm{F}\cdot\frac{\bm{v}}{v}\Bigr)}{\sqrt{1-\frac{v^2}{c^2}}}+
\hat{\iota}\frac{\frac{\bm{v}}{c}\times\bm{F}}{\sqrt{1-\frac{v^2}{c^2}}}.
\end{gather}
Consequently, transformation of the fields $\bm{E}$ and $c\bm{B}$
\begin{gather}
\bm{E}'=
\frac{\bm{E}+\frac{\bm{v}}{c}\times c\bm{B}}{\sqrt{1-\frac{v^2}{c^2}}}
+\frac{\bm{v}}{v}\Bigl(\bm{E}\cdot\frac{\bm{v}}{v}\Bigr)\Bigl(1-\frac{1}{\sqrt{1-\frac{v^2}{c^2}}}\Bigr),\\
c\bm{B}'=\frac{c\bm{B}-\frac{\bm{v}}{c}\times\bm{E}}{\sqrt{1-\frac{v^2}{c^2}}}+
\frac{\bm{v}}{v}\Bigl(c\bm{B}\cdot\frac{\bm{v}}{v}\Bigr)\Bigl(1-\frac{1}{\sqrt{1-\frac{v^2}{c^2}}}\Bigr).
\end{gather}
coincide with the well-known electric and magnetic fields boost.

So, the antisymmetric tensors of the second order can be presented as hypercomplex numbers (70). Their Lorentz transformation is provided by the same expression (56)  as for vectors.
\section{Lorentz transformations composition}
The hypercomplex representation allows us to find a simple rule for the two Lorentz's transformations composition. The
Lorentz's transformation, which is equivalent to two successive Lorentz's transformations, can be found in the following way. 
Those two transformations (the first with parameter $L'$ and
the second with parameter $L''$) can be written as the multiplications of two exponent $e^{\frac{1}{2}L'_{\alpha\beta}\sigma^{\alpha\beta}}$ and  $e^{\frac{1}{2}L''_{\alpha\beta}\sigma^{\alpha\beta}}$:
\begin{gather}
e^{\frac{1}{2}L''_{\alpha\beta}\sigma^{\alpha\beta}}e^{\frac{1}{2}L'_{\alpha\beta}\sigma^{\alpha\beta}}=\nonumber\\
=\frac{(1+\frac{1}{2}\Lambda''_{\alpha\beta}\sigma^{\alpha\beta})(1+\frac{1}{2}\Lambda'_{\alpha\beta}\sigma^{\alpha\beta})}
{\sqrt{(1+\frac{1}{2}\Lambda''^{\alpha\beta}\bullet\Lambda''_{\alpha\beta})
(1+\frac{1}{2}\Lambda'^{\alpha\beta}\bullet\Lambda'_{\alpha\beta})}}=\nonumber\\
=\frac{1-\frac{1}{2}\Lambda''^{\alpha\beta}\bullet\Lambda'_{\alpha\beta}}
{\sqrt{(1+\frac{1}{2}\Lambda''^{\alpha\beta}\bullet\Lambda''_{\alpha\beta})
(1+\frac{1}{2}\Lambda'^{\alpha\beta}\bullet\Lambda'_{\alpha\beta})}}\times\nonumber\\
\times\Bigl(1+\frac{1}{2}\frac{\Lambda'_{\alpha\beta}+\Lambda''_{\alpha\beta}+[\Lambda''_{\alpha\mu}\Lambda'^{\mu.}_{.\beta}]}
{1-\frac{1}{2}\Lambda''^{\alpha\beta}\bullet\Lambda'_{\alpha\beta}}\sigma^{\alpha\beta}\Bigr).
\end{gather}
Let us set up
\begin{equation}
\fbox{\parbox[c][11mm]{51mm}{\[
\Lambda_{\alpha\beta}=\frac{\Lambda'_{\alpha\beta}+\Lambda''_{\alpha\beta}+[\Lambda''_{\alpha\mu}\Lambda'^{\mu.}_{.\beta}]}
{1-\frac{1}{2}\Lambda''^{\alpha\beta}\bullet\Lambda'_{\alpha\beta}}
\]}}\,.
\end{equation}
It is easy to see that
\begin{gather}
\Lambda^{\alpha\beta}\bullet\Lambda_{\alpha\beta}=\frac{(1+\frac{1}{2}\Lambda''^{\alpha\beta}\bullet\Lambda''_{\alpha\beta})
(1+\frac{1}{2}\Lambda'^{\alpha\beta}\bullet\Lambda'_{\alpha\beta})}{(1-\frac{1}{2}\Lambda''^{\alpha\beta}\bullet\Lambda'_{\alpha\beta})^2}-1.
\end{gather}
Expression (79) becomes
\begin{gather}
e^{\frac{1}{2}L''_{\alpha\beta}\sigma^{\alpha\beta}}e^{\frac{1}{2}L'_{\alpha\beta}\sigma^{\alpha\beta}}
=\frac{1+\frac{1}{2}\Lambda_{\alpha\beta}\sigma^{\alpha\beta}}
{\sqrt{1+\frac{1}{2}\Lambda^{\alpha\beta}\bullet\Lambda_{\alpha\beta}}}=e^{\frac{1}{2}L_{\alpha\beta}\sigma^{\alpha\beta}}.
\end{gather}
Here
\begin{gather}
L_{\alpha\beta}=\Lambda_{\alpha\beta}\frac{L}{\tanh L},\hspace{7mm}\tanh L=\sqrt{-\frac{1}{2}\Lambda^{\alpha\beta}\bullet\Lambda_{\alpha\beta}}.
\end{gather}
Thus, the result of the two Lorentz transformations (with the parameters $\Lambda'_{\alpha\beta}$ and $\Lambda''_{\alpha\beta}$) is a 
Lorentz transformation with the parameter  $\Lambda_{\alpha\beta}$.  Expression (80) is the rule by which the Lorentz transformations 
can be converted into the parameter  composition $\Lambda_{\alpha\beta}$.

Two successive Lorentz transformations (52), the first  $e^{\frac{1}{2}l'_{kl}\sigma^{kl}}$ and the second  $e^{\frac{1}{2}l''_{kl}\sigma^{kl}}$ can be
written as
\begin{gather}
e^{\frac{1}{2}l''_{kl}\sigma^{kl}}e^{\frac{1}{2}l'_{kl}\sigma^{kl}}
=\frac{(1+\frac{1}{2}\lambda''_{kl}\sigma^{kl})(1+\frac{1}{2}\lambda'_{kl}\sigma^{kl})}
{\sqrt{(1+\frac{1}{2}\lambda''^{kl}\lambda''_{kl})
(1+\frac{1}{2}\lambda'^{kl}\lambda'_{kl})}}=\nonumber\\
=\frac{1-\frac{1}{2}\lambda''^{kl}\lambda'_{kl}}
{\sqrt{(1+\frac{1}{2}\lambda''^{kl}\lambda''_{kl})
(1+\frac{1}{2}\lambda'^{kl}\lambda'_{kl})}}\times\nonumber\\
\times\Bigl(1+\frac{1}{2}\frac{\lambda'_{kl}+\lambda''_{kl}+[\lambda''_{km}\lambda'^{m.}_{.l}]}
{1-\frac{1}{2}\lambda''^{kl}\lambda'_{kl}}\sigma^{kl}\Bigr).
\end{gather}
if we use hypercomplex numbers (16) with 8 basic units with $\hat{\iota}$-complex coefficients.

Let us set up
\begin{equation}
\fbox{\parbox[c][11mm]{51mm}{\[
\lambda_{kl}=\frac{\lambda'_{kl}+\lambda''_{kl}+[\lambda''_{km}\lambda'^{m.}_{.l}]}
{1-\frac{1}{2}\lambda''^{kl}\lambda'_{kl}}
\]}}\,.
\end{equation}
It is easy to show that
\begin{gather}
\lambda^{kl}\lambda_{kl}=\frac{(1+\frac{1}{2}\lambda''^{kl}\lambda''_{kl})
(1+\frac{1}{2}\lambda'^{kl}\lambda'_{kl})}{(1-\frac{1}{2}\lambda''^{kl}\lambda'_{kl})^2}-1.
\end{gather}
Expression (84) becomes
\begin{gather}
e^{\frac{1}{2}l''_{kl}\sigma^{kl}}e^{\frac{1}{2}l'_{kl}\sigma^{kl}}
=\frac{1+\frac{1}{2}\lambda_{kl}\sigma^{kl}}
{\sqrt{1+\frac{1}{2}\lambda^{kl}\lambda_{kl}}}=e^{\frac{1}{2}l_{kl}\sigma^{kl}}.
\end{gather}
Here
\begin{gather}
l_{kl}=\lambda_{kl}\frac{l}{\tanh l},\hspace{7mm}\tanh l=\sqrt{-\frac{1}{2}\lambda^{kl}\lambda_{kl}}.
\end{gather}
In this case, parameter composition rule (85) becomes simpler, because instead of (9), which determines the $\hat{\iota}$-complex
number $\Lambda''^{\alpha\beta}\bullet\Lambda'_{\alpha\beta}$, the regular tensor contraction $\lambda''^{kl}\lambda'_{kl}$ of
the $\hat{\iota}$-complex tensor coordinates of the tensor parameters have to be used.

The use of the $\hat{\iota}$-complex 3D vector notations (23) and (56) leads to the further composition rule simplifications:
\begin{gather}
e^{\bm{l}_2\bm{\varsigma}}e^{\bm{l}_1\bm{\varsigma}}=\frac{(1+\bm{\lambda}_2\bm{\varsigma})(1+\bm{\lambda}_1\bm{\varsigma})}
{\sqrt{(1+\bm{\lambda}_1\cdot\bm{\lambda}_1)(1+\bm{\lambda}_2\cdot\bm{\lambda}_2})}=\nonumber\\
=\frac{1-\bm{\lambda}_1\cdot\bm{\lambda}_2}{\sqrt{(1+\bm{\lambda}_1\cdot\bm{\lambda}_1)(1+\bm{\lambda}_2\cdot\bm{\lambda}_2})}
\Bigl(1+\frac{\bm{\lambda}_1+\bm{\lambda}_2+\bm{\lambda}_2\times{\lambda}_1}{1-\bm{\lambda}_1\cdot\bm{\lambda}_2}\bm{\varsigma}\Bigr),
\end{gather}
Let us set up
\begin{equation}
\fbox{\parbox[c][11mm]{40mm}{\[
\bm{\lambda}=\frac{\bm{\lambda}_1+\bm{\lambda}_2+\bm{\lambda}_2\times{\lambda}_1}{1-\bm{\lambda}_1\cdot\bm{\lambda}_2}
\]}}\,.
\end{equation}
Then
\begin{gather}
\bm{\lambda}\cdot\bm{\lambda}
=\frac{{(1+\bm{\lambda}_1\cdot\bm{\lambda}_1)(1+\bm{\lambda}_2\cdot\bm{\lambda}_2})}{(1-\bm{\lambda}_1\cdot\bm{\lambda}_2)^2}-1.
\end{gather}
Thus, expression (89) becomes
\begin{gather}
e^{\bm{l}_2\bm{\varsigma}}e^{\bm{l}_1\bm{\varsigma}}=\frac{1+\bm{\lambda}\bm{\varsigma}}
{\sqrt{1+\bm{\lambda}\cdot\bm{\lambda}}}=e^{\bm{l}\bm{\varsigma}}.
\end{gather}
Here
\begin{gather}
\bm{l}=\frac{l}{\tanh l}\bm{\lambda},\hspace{4mm}\tanh l=\sqrt{-\bm{\lambda}\cdot\bm{\lambda}}.
\end{gather}
As we can see, the use of the composition rule (90) in this case requires a regular dot-product and cross-product of  $\hat{\iota}$-complex
3D vectors only.

Composition rule (90) coincides with Fedorov's rule (1), but   $\bm{\lambda}$ in (90) is an $\hat{\iota}$-complex 3D coefficient of the hypercomplex
numbers $\bm{\lambda}\bm{\varsigma}$ (i.e. $4\times 4$ matrices). Lorentz transformations are performed  by $e^{\pm\bm{\lambda}\bm{\varsigma}}$
exponents which are usual for the relativistic quantum mechanics approach.
\section{Boost composition}
Let us treat boost composition, which is a particular but important case of the Lorentz transformation composition.  Assume that we have two boosts.
The first with the parameter
\begin{equation}
\bm{\lambda}_1(\hat{\iota}\bm{b}_1)=\hat{\iota}\frac{\bm{b}_1}{b_1}\tanh b_1=\hat{\iota}\frac{\bm{v}_1}{c}\frac{1}{1+\sqrt{1-\frac{v_1^2}{c^2}}}
\end{equation}
and the second with the parameter
\begin{equation}
\bm{\lambda}_2(\hat{\iota}\bm{b}_2)=\hat{\iota}\frac{\bm{b}_2}{b_2}\tanh b_2=\hat{\iota}\frac{\bm{v}_2}{c}\frac{1}{1+\sqrt{1-\frac{v_2^2}{c^2}}}.
\end{equation}
 According to (90), parameter transformation, which is the result of the two successive above transformations is
\begin{gather}
\bm{\lambda}(\bm{l})=\frac{\bm{l}}{l}\tanh l=\nonumber\\=
\frac{\hat{\iota}(\frac{\bm{b}_1}{b_1}\tanh b_1+\frac{\bm{b}_2}{b_2}\tanh b_2)+(\frac{\bm{b}_1}{b_1}\times\frac{\bm{b}_2}{b_2})\tanh b_1\tanh b_2}
{1+(\frac{\bm{b}_1}{b_1}\cdot\frac{\bm{b}_2}{b_2})\tanh b_1\tanh b_2}.
\end{gather}
Because $\bm{\lambda}(\bm{l})$ is an $\hat{\iota}$-complex number, the resulting transformation is neither a boost nor rotation.
As any Lorentz transformation it can be represented as the result of rotation and boost:
\begin{equation}
e^{\bm{l}\bm{\varsigma}}=e^{\hat{\iota}\bm{b}\bm{\varsigma}}e^{\bm{r}\bm{\varsigma}}
\end{equation}
or as the result of boost and rotation:
\begin{equation}
e^{\bm{l}\bm{\varsigma}}=e^{\bm{r'}\bm{\varsigma}}e^{\hat{\iota}\bm{b}'\bm{\varsigma}}.
\end{equation}
Two successive boosts $e^{\hat{\iota}2\bm{b}\bm{\varsigma}}$ and $e^{\hat{\iota}2\bm{b}'\bm{\varsigma}}$ can be found by
multiplying both parts of equations (97) and (98) by their Hermitian conjugate expressions
\begin{gather}
(e^{\bm{l}\bm{\varsigma}})^\dag=e^{-\breve{\bm{l}}\bm{\varsigma}}=e^{-\bm{r}\bm{\varsigma}}e^{\hat{\iota}\bm{b}\bm{\varsigma}},\\
(e^{\bm{l}\bm{\varsigma}})^\dag=e^{-\breve{\bm{l}}\bm{\varsigma}}=e^{\hat{\iota}\bm{b}'\bm{\varsigma}}e^{-\bm{r}'\bm{\varsigma}}.
\end{gather}
Let us multiply equation (97)  from the right by $e^{-\breve{\bm{l}}\bm{\varsigma}}$ and equation (98) from the left by $e^{-\breve{\bm{l}}\bm{\varsigma}}$:
\begin{gather}
e^{\bm{l}\bm{\varsigma}}e^{-\breve{\bm{l}}\bm{\varsigma}}=
e^{\hat{\iota}\bm{b}\bm{\varsigma}}e^{\bm{r}\bm{\varsigma}}e^{-\bm{r}\bm{\varsigma}}e^{\hat{\iota}\bm{b}\bm{\varsigma}}=
e^{\hat{\iota}2\bm{b}\bm{\varsigma}},\\
e^{-\breve{\bm{l}}\bm{\varsigma}}e^{\bm{l}\bm{\varsigma}}=
e^{\hat{\iota}\bm{b}'\bm{\varsigma}}e^{-\bm{r}'\bm{\varsigma}}e^{\bm{r}'\bm{\varsigma}}e^{\hat{\iota}\bm{b}'\bm{\varsigma}}=
e^{\hat{\iota}2\bm{b}'\bm{\varsigma}}.
\end{gather}
Transformation parameters $e^{\bm{l}\bm{\varsigma}}e^{-\breve{\bm{l}}\bm{\varsigma}}=e^{\hat{\iota}2\bm{b}\bm{\varsigma}}$ and
$e^{-\breve{\bm{l}}\bm{\varsigma}}e^{\bm{l}\bm{\varsigma}}=e^{\hat{\iota}2\bm{b}'\bm{\varsigma}}$,  according to the (90), are 
\begin{gather}
\bm{\lambda}(\hat{\iota}2\bm{b})=\hat{\iota}\frac{\bm{b}}{b}\tanh 2b=\hat{\iota}\frac{\bm{v}}{c}=
\frac{\bm{\lambda}(\bm{l})-\breve{\bm{\lambda}}(\bm{l})-\bm{\lambda}(\bm{l})\!\times\!\breve{\bm{\lambda}}(\bm{l})}
{1+\bm{\lambda}(\bm{l})\cdot\breve{\bm{\lambda}}(\bm{l})}=\nonumber\\=
2\frac{\bm{\lambda}_1(1+\bm{\lambda}_2\cdot\bm{\lambda}_2)+\bm{\lambda}_2(1-\bm{\lambda}_1\cdot\bm{\lambda}_1)-2\bm{\lambda}_2(\bm{\lambda}_1\cdot\bm{\lambda}_2)}
{(1-\bm{\lambda}_1\cdot\bm{\lambda}_1)(1-\bm{\lambda}_2\cdot\bm{\lambda}_2)-4(\bm{\lambda}_1\cdot\bm{\lambda}_2)}=\nonumber\\=
\hat{\iota}\frac{\frac{\bm{v}_1}{c}\sqrt{1-\frac{v_2^2}{c^2}}+\frac{\bm{v}_2}{c}\Bigl[1+\frac{\bm{v}_1}{c}\!\cdot\!\frac{\bm{v}_2}{c}
\bigl({1+\sqrt{1-\frac{v_2^2}{c^2}}}\bigr)^{-1}\Bigr]}{1+\frac{\bm{v}_1}{c}\cdot\frac{\bm{v}_2}{c}},\\
\bm{\lambda}(\hat{\iota}2\bm{b}')\!=\!\hat{\iota}\frac{\bm{b}'}{b'}\tanh 2b'=\hat{\iota}\frac{\bm{v}'}{c}=
\frac{\bm{\lambda}(\bm{l})\!-\!\breve{\bm{\lambda}}(\bm{l})\!-\!\breve{\bm{\lambda}}(\bm{l})\!\times\!\bm{\lambda}(\bm{l})}
{1+\bm{\lambda}(\bm{l})\cdot\breve{\bm{\lambda}}(\bm{l})}=\nonumber\\=
2\frac{\bm{\lambda}_2(1+\bm{\lambda}_1\cdot\bm{\lambda}_1)+\bm{\lambda}_1(1-\bm{\lambda}_2\cdot\bm{\lambda}_2)-2\bm{\lambda}_1(\bm{\lambda}_1\cdot\bm{\lambda}_2)}
{(1-\bm{\lambda}_1\cdot\bm{\lambda}_1)(1-\bm{\lambda}_2\cdot\bm{\lambda}_2)-4(\bm{\lambda}_1\cdot\bm{\lambda}_2)}=\nonumber\\=
\hat{\iota}\frac{\frac{\bm{v}_2}{c}\sqrt{1-\frac{v_1^2}{c^2}}+\frac{\bm{v}_1}{c}\Bigl[1+\frac{\bm{v}_1}{c}\!\cdot\!\frac{\bm{v}_2}{c}
\bigl({1+\sqrt{1-\frac{v_1^2}{c^2}}}\bigr)^{-1}\Bigr]}{1+\frac{\bm{v}_1}{c}\cdot\frac{\bm{v}_2}{c}}.
\end{gather}
Here $\bm{v}$ and $\bm{v}'$ are boost velocities represented by the exponents $e^{\hat{\iota}\bm{b}\bm{\varsigma}}$ and $e^{\hat{\iota}\bm{b}'\bm{\varsigma}}$.
These two velocities are equal by magnitude 
\begin{gather}
\frac{{v}}{c}=\frac{{v}'}{c}=\sqrt{1-\frac{(1-\frac{v_1^2}{c^2})(1-\frac{v_2^2}{c^2})}{(1+\frac{\bm{v}_1}{c}\frac{\bm{v}_2}{c})^2}},
\end{gather}
but have different directions.

Expressions (103)-(104) determine, in particular, the relativistic rule of velocity addition.  Let us assume that  reference frame 
$S_{-\bm{v}_1}$ is moving with velocity $-\bm{v}_1$ relative to reference frame  $S$. Evidently, a particle which 
 is at rest in reference frame $S$ has velocity $\bm{v}_1$ in the reference frame $S_{-\bm{v}_1}$. 
  Assume that we also have reference frame $S_{-\bm{v}_2}$,
 which is moving with velocity $-\bm{v}_2$ relative to the reference frame  $S_{-\bm{v}_1}$.  The same particle in this
 reference frame has a  velocity which is opposite to the velocity of the reference frame $S_{-\bm{v}_2}$  relative to
 the reference frame  $S$. We represent the transition from  $S$ to $S_{-\bm{v}_2}$ (using (97)) as rotation and
 successive boost.  Rotation does not change the  velocity of the particle, which is at rest
in reference frame $S$, but it acquires velocity (103) due to boost. Thus, a particle which has velocity   $\bm{v}_1$ in the
reference frame $S_{-\bm{v}_1}$ has velocity
\begin{gather}
\frac{\bm{v}}{c}=\frac{\frac{\bm{v}_1}{c}\sqrt{1-\frac{v_2^2}{c^2}}+\frac{\bm{v}_2}{c}\Bigl[1+\frac{\bm{v}_1}{c}\!\cdot\!\frac{\bm{v}_2}{c}
\bigl({1+\sqrt{1-\frac{v_2^2}{c^2}}}\bigr)^{-1}\Bigr]}{1+\frac{\bm{v}_1}{c}\cdot\frac{\bm{v}_2}{c}}=\nonumber\\
=\frac{\frac{\bm{v}_1}{c}+\frac{\bm{v}_2}{c}+\frac{\bm{v}_2}{c}\times(\frac{\bm{v}_2}{c}\times\frac{\bm{v}_1}{c})(1+\sqrt{1-\frac{v_2^2}{c^2}})^{-1}}
{1+\frac{\bm{v}_1}{c}\cdot\frac{\bm{v}_2}{c}}
\end{gather}
in the reference frame  $S_{-\bm{v}_2}$, which is moving with velocity $-\bm{v}_2$ relative to the reference frame  $S_{-\bm{v}_1}$.
This is a well-known relativistic velocity addition rule  \cite{fo}.

We can represent a transformation from reference frame  $S$ to  reference frame  $S_{-\bm{v}_2}$ as a boost and successive rotation.
After  a boost, the particle has velocity (104), but this velocity differs from the final velocity by rotation  $e^{\bm{r}'\bm{\varsigma}}$.

Note also, that a particle which has  velocity $\bm{v}_2$ in the reference frame $S_{-\bm{v}_2}$, has velocity (104) in the reference frame $S_{-\bm{v}_1}$.
Reference frame $S_{-\bm{v}_1}$ is moving with velocity  $-\bm{v}_1$ relative to  $S_{-\bm{v}_2}$.
Thus, a switch of the particle velocity and the reference frame velocity is a rotation of the calculated velocity represented by the exponent  $e^{\bm{r}'\bm{\varsigma}}$.

Let us calculate the boost parameters represented in (97)-(98) by exponents $e^{\hat{\iota}\bm{b}\bm{\varsigma}}$ and $e^{\hat{\iota}\bm{b}'\bm{\varsigma}}$ and
rotation parameters represented by exponents $e^{\bm{r}\bm{\varsigma}}$ и $e^{\bm{r}'\bm{\varsigma}}$. Parameters $\bm{\lambda}(\hat{\iota}\bm{b})$ and
$\bm{\lambda}(\hat{\iota}\bm{b}')$ of the boosts $e^{\hat{\iota}\bm{b}\bm{\varsigma}}$ and $e^{\hat{\iota}\bm{b}'\bm{\varsigma}}$ differ from the parameters
$\bm{\lambda}(\hat{\iota}2\bm{b})$ and $\bm{\lambda}(\hat{\iota}2\bm{b}')$ of the boosts $e^{\hat{\iota}2\bm{b}\bm{\varsigma}}$ and
$e^{\hat{\iota}2\bm{b}'\bm{\varsigma}}$ by multiplier
\begin{gather}
\left.
\begin{array}{lcc}
\frac{\tanh b}{\tanh 2b}&=&\Bigl[1+\sqrt{1+\bm{\lambda}(\hat{\iota}2\bm{b})\cdot\bm{\lambda}(\hat{\iota}2\bm{b})}\Bigr]^{-1}\\
\frac{\tanh b'}{\tanh 2b'}&=&\Bigl[1+\sqrt{1+\bm{\lambda}(\hat{\iota}2\bm{b}')\cdot\bm{\lambda}(\hat{\iota}2\bm{b}')}\Bigr]^{-1}
\end{array}
\right\}=\nonumber\\
=\frac{1}{2}\frac{(1-\bm{\lambda}_1\cdot\bm{\lambda}_1)(1-\bm{\lambda}_2\cdot\bm{\lambda}_2)-4\bm{\lambda}_1\cdot\bm{\lambda}_2}
{1+(\bm{\lambda}_1\cdot\bm{\lambda}_1)(\bm{\lambda}_2\cdot\bm{\lambda}_2)-2\bm{\lambda}_1\cdot\bm{\lambda}_2}.
\end{gather}
Thus, boost parameters are represented by the exponents $e^{\hat{\iota}\bm{b}\bm{\varsigma}}$ and $e^{\hat{\iota}\bm{b}'\bm{\varsigma}}$ are
equal to 
\begin{gather}
\bm{\lambda}(\hat{\iota}\bm{b})=\hat{\iota}\frac{\bm{b}}{b}\tanh b=\nonumber\\=
\frac{\bm{\lambda}_1(1+\bm{\lambda}_2\cdot\bm{\lambda}_2)+\bm{\lambda}_2[1-\bm{\lambda}_1\cdot\bm{\lambda}_1-2\bm{\lambda}_1\cdot\bm{\lambda}_2]}
{1+(\bm{\lambda}_1\cdot\bm{\lambda}_1)(\bm{\lambda}_2\cdot\bm{\lambda}_2)-2\bm{\lambda}_1\cdot\bm{\lambda}_2},\\
\bm{\lambda}(\hat{\iota}\bm{b}')=\hat{\iota}\frac{\bm{b}'}{b'}\tanh b'=\nonumber\\=
\frac{\bm{\lambda}_2(1+\bm{\lambda}_1\cdot\bm{\lambda}_1)+\bm{\lambda}_1[1-\bm{\lambda}_2\cdot\bm{\lambda}_2-2\bm{\lambda}_1\cdot\bm{\lambda}_2]}
{1+(\bm{\lambda}_1\cdot\bm{\lambda}_1)(\bm{\lambda}_2\cdot\bm{\lambda}_2)-2\bm{\lambda}_1\cdot\bm{\lambda}_2}.
\end{gather}
With the help of parameters  $\bm{\lambda}(\hat{\iota}\bm{b})$ and $\bm{\lambda}(\hat{\iota}\bm{b}')$ we can find rotation parameters represented in  (97)-(98)
by exponents $e^{\bm{r}\bm{\varsigma}}$ and $e^{\bm{r}' \bm{\varsigma}}$.
Let us multiply (97) by $e^{-\hat{\iota}\bm{b}\bm{\varsigma}}$ from the left, multiply (98) by  $e^{-\hat{\iota}\bm{b}'\bm{\varsigma}}$ from the right and use
 (90):
\begin{gather}
\left.
\begin{array}{lr}
\lambda(\bm{r})\!=\!\frac{\bm{r}}{r}\tanh r\!=\!\frac{\lambda(\bm{l})-\lambda(\hat{\iota}\bm{b})-\lambda(\hat{\iota}\bm{b})\!\times\!\lambda(\bm{l})}
{1+\lambda(\bm{l})\cdot\lambda(\hat{\iota}\bm{b})}\\
\lambda(\bm{r}')\!=\!\frac{\bm{r}'}{r'}\tanh r'\!=\!\frac{\lambda(\bm{l})-\lambda(\hat{\iota}\bm{b}')-\lambda(\bm{l})\!\times\!\lambda(\hat{\iota}\bm{b}')}
{1+\lambda(\bm{l})\cdot\lambda(\hat{\iota}\bm{b}')}
\end{array}\!\!\!
\right\}\!=\!\frac{\lambda_2\times\lambda_1}{1-\lambda_1\!\cdot\!\lambda_2}=\nonumber\\
=\frac{\bm{v}_1\times\bm{v}_2}{c^2\bigl(1+\sqrt{1-\frac{v_1^2}{c^2}}\bigr)\bigl(1+\sqrt{1-\frac{v_2^2}{c^2}}\bigr)+\bm{v}_1\cdot\bm{v}_2}.
\end{gather}
As we can see rotation is the same in both cases.

In particular, expression (109) represents Thomas rotation \cite{t}. In this case, in the reference frame  $S$ during a small time interval, the  
$\delta t$  particle
changes its velocity from $\bm{v}$ to $\bm{v}+\delta\bm{v}$.  This velocity change is equivalent to the velocity change of the reference frame in the
opposite direction.  We can represent the velocity change  in two steps. First, reference frame $S$ gets the velocity $\bm{v}$.  In this reference
frame, the particle stays at rest.  Then reference frame $S$ gets the velocity $-(\bm{v}+\delta\bm{v})$.  The particle in this reference frame obtains the 
velocity $\bm{v}+\delta\bm{v}$.  These two reference frame  $S$  boosts with velocities $\bm{v}_1=\bm{v}$ and $\bm{v}_2=-(\bm{v}+\delta\bm{v})$ can
be substituted by one boost with velocity  $\bm{v}'$ (104) and a successive rotation  $e^{\bm{r}'\bm{\varsigma}}$ with the parameter  $\bm{\lambda}(\bm{r}')$ (109).
As a first approximation by $\delta\bm{v}$ the parameter is
\begin{gather}
\bm{\lambda}(\bm{r}')\approx\frac{\bm{r}'}{r'}\frac{\delta\varphi}{2}=
-\frac{\bm{v}\times\delta\bm{v}}{c^2\bigl(1+\sqrt{1-\frac{v^2}{c^2}}\bigr)^2-v^2}=\nonumber\\=
\frac{\delta\bm{v}\times\bm{v}}{2v^2}\Bigl(\frac{1}{\sqrt{1-\frac{v^2}{c^2}}}-1\Bigr).
\end{gather}
Consequently, the reference frame connected to the particle has precession relative to $S$ in the opposite direction with angular velocity
\begin{gather}
\bm{\omega}=\frac{\bm{r}'}{r'}\frac{\delta\varphi}{\delta t}=\frac{\bm{v}}{v^2}\times\frac{\delta\bm{v}}{\delta t}
\Bigl(\frac{1}{\sqrt{1-\frac{v^2}{c^2}}}-1\Bigr)=\nonumber\\=
\frac{\bm{v}\times\bm{a}}{v^2}\Bigl(\frac{1}{\sqrt{1-\frac{v^2}{c^2}}}-1\Bigr),
\end{gather}
where $\bm{a}=\delta\bm{v}/\delta t$ --- particle acceleration.  This is the Thomas precession.


\section{Spin, Pauli-Lubanski pseudo-vector and Wigner little group }
The representation of the Lorentz transformation by hypercomplex numbers clarifies the connection of spin to the Pauli-Lubanski 
pseudo-vector and the Wigner little group.

It is well known \cite{bd} that fermion, which is at rest and has spin  $\pm\hbar/2$ projections on the  $\bm{n}$ axis, is described
by bispinors $\psi_{\pm}(\bm{n})$.  These bispinors are proper $\frac{\hbar}{2}i\bm{n}\bm{\varsigma}$ operator vectors:
\begin{equation}
\frac{\hbar}{2}i\bm{n}\bm{\varsigma}\psi_{\pm}(\bm{n})=\pm\frac{\hbar}{2}\psi_{\pm}(\bm{n}).
\end{equation}
Here $\bm{n}\bm{\varsigma}=n_{23}\sigma^{23}+n_{31}\sigma^{31}+n_{12}\sigma^{12}$, $\bm{n}\bm{\varsigma}\bm{n}\bm{\varsigma}=-1$.
Normalized bispinors $\psi_{\pm}(\bm{n})$ for a fermion with positive energy can be represented as
\begin{gather}
\psi_{\pm}(\bm{n})=e^{\bm{r}\bm{\varsigma}}\psi(\bm{e}_z)=
\frac{1-(\pm \bm{n}\bm{\varsigma})(\bm{e}_z\bm{\varsigma})}{\sqrt{2(1\pm n_z)}}\psi(\bm{e}_z)=\nonumber\\=
\frac{1\pm i\bm{n}\bm{\varsigma}}{\sqrt{2(1\pm n_z)}}\left[\begin{array}{c}
  1\\
  0\\
  0\\
  0
\end{array}\right].
\end{gather}
I.e. we can represent them as a result of bispinor rotation
\begin{gather}
\psi(\bm{e}_z)=\left[\begin{array}{c}
  1\\
  0\\
  0\\
  0
\end{array}\right],
\end{gather}
which describes a fermion with spin projection along the $z$ axis equals  $\hbar/2$ .
Bispinor $\psi(\bm{e}_z)$ is the proper vector of the operator $\frac{\hbar}{2}i\bm{e}_z\bm{\varsigma}=\frac{\hbar}{2}i\sigma^{12}$:
\begin{gather}
\frac{\hbar}{2}i\bm{e}_z\bm{\varsigma}\psi(\bm{e}_z)=\frac{\hbar}{2}i\sigma^{12}\psi(\bm{e}_z)=\frac{\hbar}{2}\psi(\bm{e}_z).
\end{gather}
Exponent
\begin{gather}
e^{\bm{r}\bm{\varsigma}}=\cos r+\frac{\bm{r}\bm{\varsigma}}{r}\sin r=\nonumber\\=
\sqrt{\frac{1+\bm{e}_z\cdot\bm{n}}{2}}+\frac{(\bm{e}_z\times\bm{n})\bm{\varsigma}}{\sqrt{1-(\bm{e}_z\cdot\bm{n})^2}}\sqrt{\frac{1-\bm{e}_z\cdot\bm{n}}{2}}=
\nonumber\\=\frac{1-\bm{n}\bm{\varsigma\,\bm{e}_z\bm{\varsigma}}}{\sqrt{2(1+\bm{e}_z\cdot\bm{n})}}.
\end{gather}
represents rotation, which aligns unit vector $\bm{e}_z$ along unit vector $\bm{n}$:
\begin{gather}
e^{\bm{r}\bm{\varsigma}}\bm{e}_z\bm{\gamma}e^{-\bm{r}\bm{\varsigma}}=\bm{n}\bm{\gamma}.
\end{gather}

Bispinors $\psi_{\pm}(\bm{n})$ describe fermions at rest.  There is no change of their characteristics after reference frame rotation about $\bm{n}$ axis.
After this rotation, bispinors $\psi_{\pm}(\bm{n})$ get unitary multipliers only, but their structure remains unchanged:
\begin{equation}
\psi'_{\pm}(\bm{n})=e^{\bm{n}\bm{\varsigma}\frac{\varphi}{2}}\psi_{\pm}(\bm{n})=e^{\mp i\frac{\varphi}{2}}\psi_{\pm}(\bm{n}).
\end{equation}
There are more complex changes of the bispinors $\psi_{\pm}(\bm{n})$ after rotation about another axis.
That is, fermions at rest, which are described by bispinors $\psi_{\pm}(\bm{n})$ obey the symmetry with respect to the group transformation rotation
about the $\bm{n}$ axis. The infinitesimal operators $\frac{\varphi}{2}\bm{n}\bm{\varsigma}$ of these transformations coincide (within the multiplier)
with operator $\frac{\hbar}{2}i\bm{n}\bm{\varsigma}$. For this operator, bispinors $\psi_{\pm}(\bm{n})$ are the proper vectors.
This approach sheds light on the connection between the proper operator $\frac{\hbar}{2}i\bm{n}\bm{\varsigma}$ bispinors and the
symmetry transformations  $e^{\bm{n}\bm{\varsigma}\frac{\varphi}{2}}$ for which operator  $\frac{\hbar}{2}i\bm{n}\bm{\varsigma}$ is
 infinitesimally small (within the multiplier).

Note also, that rotation $e^{\bm{n}\bm{\varsigma}\frac{\varphi}{2}}$, as any spatial rotation, does not change linear momentum
of the fermion at rest.

In order to turn to the moving fermions it is enough to move from the proper reference frame into the frame $S_{-\bm{v}}$,
which is moving with velocity  $-\bm{v}$ relative to it.   In the reference frame  $S_{-\bm{v}}$
fermion has 4D impulse $p_0=mc/\sqrt{1-\frac{v^2}{c^2}}$, $\bm{p}=m\bm{v}/\sqrt{1-\frac{v^2}{c^2}}$.
In order to come up with equation (112) in the reference frame $S_{-\bm{v}}$, apply boost operator   $e^{\hat{\iota}\bm{b}\bm{\varsigma}}$ to both parts
of it:
\begin{gather}
\frac{\hbar}{2}\underbrace{e^{\hat{\iota}\bm{b}\bm{\varsigma}}i\bm{n}\bm{\varsigma}e^{-\hat{\iota}\bm{b}\bm{\varsigma}}}_{\bm{S}\bm{\varsigma}}
e^{\hat{\iota}\bm{b}\bm{\varsigma}}\psi_{\pm}(\bm{n})=\frac{\hbar}{2}i\bm{S}\bm{\varsigma}e^{\hat{\iota}\bm{b}\bm{\varsigma}}\psi_{\pm}(\bm{n})=\nonumber\\=
\pm\frac{\hbar}{2}e^{\hat{\iota}\bm{b}\bm{\varsigma}}\psi_{\pm}(\bm{n}).
\end{gather}
Here $\frac{\bm{b}}{|\bm{b}|}=-\frac{\bm{v}}{|\bm{v}|}$, $\tanh 2|\bm{b}|=\frac{v}{c}$.  As we can see bispinors
\begin{gather}
e^{\hat{\iota}\bm{b}\bm{\varsigma}}\psi_{\pm}(\bm{n})=\nonumber\\=
\underbrace{\sqrt{\frac{p_0+mc}{2mc}}\Bigl(1-\frac{\hat{\iota}\bm{p}\bm{\varsigma}}{p_0+mc}\Bigr)}_{e^{\hat{\iota}\bm{b}\bm{\varsigma}}}
\frac{1\pm i\bm{n}\bm{\varsigma}}{\sqrt{2(1\pm n_z)}}\psi(\bm{e}_z)
\end{gather}
are the proper bispinors of the operator
\begin{gather}
\frac{\hbar}{2}i\bm{S}\bm{\varsigma}=\frac{\hbar}{2}e^{\hat{\iota}\bm{b}\bm{\varsigma}}i\bm{n}\bm{\varsigma}e^{-\hat{\iota}\bm{b}\bm{\varsigma}}=
\nonumber\\=\frac{\hbar}{2}i
\underbrace{\frac{1}{mc}\Bigl(p_0\bm{n}-\frac{\bm{n}\cdot\bm{p}}{p_0+mc}\bm{p}+\hat{\iota}\bm{n}\times\bm{p}\Bigr)}_{\bm{S}}\bm{\varsigma}.
\end{gather}
Obviously, bispinor $e^{\hat{\iota}\bm{b}\bm{\varsigma}}\psi_{\pm}(\bm{n})$ represents bispinor  $\psi_{\pm}(\bm{n})$ in the reference frame
$S_{-\bm{v}}$.  Likewise, operator $\frac{\hbar}{2}i\bm{S}\bm{\varsigma}$ represents operator $\frac{\hbar}{2}i\bm{n}\bm{\varsigma}$ in the
reference frame $S_{-\bm{v}}$. Operator $\frac{\hbar}{2}i\bm{S}\bm{\varsigma}$ can be written in the usual form for moment operators. 
We can represent
tensor $\frac{\hbar}{2}i\bm{n}\bm{\varsigma}$ as the product of space-like pseudo-vector $n_\alpha\pi^\alpha=0\pi^0-\bm{n}\bm{\pi}$ and time-like
linear momentum $p_\alpha\gamma^\alpha=mc\gamma^0-\bm{0}\bm{\gamma}$ in the proper fermion reference frame:
\begin{equation}
\frac{\hbar}{2}i\bm{n}\bm{\varsigma}=\frac{\hbar}{2}i\frac{1}{mc}
\underbrace{\bm{n}\bm{\pi}mc\gamma^0}_{mc\bm{n}\bm{\varsigma}}=\frac{\hbar}{2}i\frac{(-1)}{mc}n_\alpha\pi^\alpha mc\gamma^0.
\end{equation}
$n_\alpha\pi^\alpha=-\bm{n}\bm{\pi}$ is a Pauli-Lubanski pseudo-vector in its proper fermion reference frame  \cite{l}. Its direction coincides 
with the rotation symmetry axis of the fermion at rest.
Using (121)-(122), for the operator $\frac{\hbar}{2}i\bm{S}\bm{\varsigma}$ in the frame $S_{-\bm{v}}$ we obtain
\begin{gather}
\frac{\hbar}{2}i\bm{S}\bm{\varsigma}
=\frac{\hbar}{2}i\frac{1}{mc}\underbrace{e^{\hat{\iota}\bm{b}\bm{\varsigma}}\bm{n}\bm{\pi}e^{-\hat{\iota}\bm{b}\bm{\varsigma}}}_{-s_\alpha\pi^\alpha}
\underbrace{e^{\hat{\iota}\bm{b}\bm{\varsigma}}mc\gamma^0e^{-\hat{\iota}\bm{b}\bm{\varsigma}}}_{p_\beta\gamma^\beta}=\nonumber\\=
\frac{\hbar}{2}i\frac{(-1)}{mc}s_\alpha\pi^\alpha p_\beta\gamma^\beta=
\frac{\hbar}{2}i\frac{1}{2}\underbrace{\frac{1}{mc}[s_\alpha p_\beta]^\diamond}_{S_{\alpha\beta}}\sigma^{\alpha\beta}=\nonumber\\=
\frac{\hbar}{2}i\frac{1}{2}S_{\alpha\beta}\sigma^{\alpha\beta}=
\frac{\hbar}{2}i\underbrace{\frac{1}{mc}[(p_0\bm{s}-s_0\bm{p})+\hat{\iota}\bm{s}\times\bm{p}]}_{\bm{S}}\bm{\varsigma}.
\end{gather}
Here
\begin{gather}
s_\alpha\pi^\alpha=e^{\hat{\iota}\bm{b}\bm{\varsigma}}n_\alpha\pi^\alpha e^{-\hat{\iota}\bm{b}\bm{\varsigma}}=s_0\pi^0-\bm{s}\bm{\pi}
\end{gather}
--- Pauli-Lubanski pseudo-vector in reference frame $S_{-\bm{v}}$,
\begin{gather}
s_0=\frac{\bm{n}\cdot\bm{p}}{mc},\hspace{7mm}\bm{s}=\bm{n}+
\frac{\bm{n}\cdot\bm{p}}{mc}\frac{\bm{p}}{p_0+mc},\\
\bm{S}=\frac{1}{mc}[(p_0\bm{s}-s_0\bm{p})+\hat{\iota}\bm{s}\times\bm{p}],\\
S_{\alpha\beta}=\frac{1}{mc}[s_\alpha p_\beta]^\diamond=\nonumber\\=
\frac{1}{mc}\frac{1}{2}\varepsilon_{\alpha\beta\mu\nu}(s^\mu p^\nu-s^\nu p^\mu)=
\frac{1}{mc}\varepsilon_{\alpha\beta\mu\nu}s^\mu p^\nu.
\end{gather}
It is easy to show that $s_\alpha p^\alpha =s_0p^0-\bm{s}\cdot\bm{p}=0$, and
\begin{gather}
s_\alpha\pi^\alpha=\frac{1}{mc}\bm{S}\bm{\varsigma}p_\beta\gamma^\beta\hspace{4mm}\mbox{or}\hspace{4mm}
s^\alpha=\frac{\varepsilon^{\mu\nu\beta\alpha}}{2mc}S_{\mu\nu}p_\beta.
\end{gather}
Usually this expression is used for the definition of the Pauli-Lubanski pseudo-vector.

From the expression (127) for $S_{\alpha\beta}$ we can see that the structure of the tensor of the 4D momentum in classical mechanics
\begin{gather}
M_{\alpha\beta}=x_\alpha p_\beta-x_\beta p_\alpha
\end{gather}
is similar to the structure of the tensor
 \begin{gather}
\frac{\hbar}{mc}[s_\alpha p_\beta]=\frac{\hbar}{mc}s_\alpha p_\beta-\frac{\hbar}{mc}s_\beta p_\alpha,
\end{gather}
dual to the $\hbar S_{\alpha\beta}$. In this sense, pseudo-vector  $s_\alpha\hbar/mc$ is the inner (proper) space-like pseudo-vector,
which has Compton length  $\hbar/mc$ and direction along the rotational symmetry axis of of the fermion at rest.

Tensor of 4D momentum $M_{\alpha\beta}$ is similar not to the $\hbar S_{\alpha\beta}$ tensor, but to $\frac{\hbar}{mc}[s_\alpha p_\beta]$,
which is dual to it. The reason is that $x_\alpha$ is a 4D-vector, and $s_\alpha$ is a 4D  pseudo-vector.

The tensors $\hbar S_{\alpha\beta}$ and $M_{\alpha\beta}$ duality is reflected in the similarity of the structure of the 3D pseudo-vector part of the tensor
$\hbar S_{\alpha\beta}$
\begin{gather}
\frac{\hbar}{mc}(p_0\bm{s}-s_0\bm{p})\bm{\varsigma}
\end{gather}
to the structure of the 3D vector part of the tensor $M_{\alpha\beta}$
\begin{gather}
p_0\bm{x}-x_0\bm{p}.
\end{gather}
Although they have similar structures, part (131) contributes to the angular momentum conservation law while part (132) contributes to the law of  the center mass motion. And vice versa, while the structure of the 3D vector part of the tensor $\hbar S_{\alpha\beta}$
\begin{gather}
\hat{\iota}\frac{\hbar}{mc}(\bm{s}\times\bm{p})\bm{\varsigma}
\end{gather}
is similar to the structure of the 3D pseudo-vector part of the tensor $M_{\alpha\beta}$
\begin{gather}
\bm{x}\times\bm{p},
\end{gather}
the (133) part contributes to the law of the center mass motion while the part (134) contributes to the angular momentum conservation law.
Again, it is because $\bm{x}$ is a vector and $\bm{s}$ is a pseudo-vector.

Note also, that pseudo-vector $s^\alpha$ is equal to the bilinear combination  $i\bar{\psi}\pi^\alpha\psi$ of the bispinors (113)
\begin{gather}
s_\alpha=i\bar{\psi}\pi^\alpha\psi,
\end{gather}
Also tensor $S^{\alpha\beta}$ is bilinear combination  $i\bar{\psi}\sigma^{\alpha\beta}\psi$:
\begin{gather}
S^{\alpha\beta}=i\bar{\psi}\sigma^{\alpha\beta}\psi.
\end{gather}

Operator $\frac{\hbar}{2}i\bm{S}\bm{\varsigma}$, similar to  $\frac{\hbar}{2}i\bm{n}\bm{\varsigma}$ (in its proper fermion reference frame),
coincides within the multiplier with an infinitesimal operator of the Lorentz transformation which results in the multiplication of bispinor
$e^{\hat{\iota}\bm{b}\bm{\varsigma}}\psi_{\pm}$  by $e^{\mp\frac{\varphi}{2}}$.
In order to prove this, let us write (118) in the reference frame $S_{-\bm{v}}$:
\begin{equation}
\underbrace{e^{\hat{\iota}\bm{b}\bm{\varsigma}}e^{\frac{\varphi}{2}\bm{n}\bm{\varsigma}}e^{-\hat{\iota}\bm{b}\bm{\varsigma}}}_
{e^{\frac{\varphi}{2}\bm{S}\bm{\varsigma}}}
e^{\hat{\iota}\bm{b}\bm{\varsigma}}\psi_{\pm}=e^{\mp i\frac{\varphi}{2}}e^{\hat{\iota}\bm{b}\bm{\varsigma}}\psi_{\pm}.
\end{equation}
The product $e^{\hat{\iota}\bm{b}\bm{\varsigma}}e^{\frac{\varphi}{2}\bm{n}\bm{\varsigma}}e^{-\hat{\iota}\bm{b}\bm{\varsigma}}$
can be written as
\begin{gather}
e^{\hat{\iota}\bm{b}\bm{\varsigma}}e^{\frac{\varphi}{2}\bm{n}\bm{\varsigma}}e^{-\hat{\iota}\bm{b}\bm{\varsigma}}=
e^{\hat{\iota}\bm{b}\bm{\varsigma}}(\cos \frac{\varphi}{2}+\bm{n}\bm{\varsigma}\sin \frac{\varphi}{2})e^{-\hat{\iota}\bm{b}\bm{\varsigma}_2}=\nonumber\\
=\cos \frac{\varphi}{2}+\underbrace{e^{\hat{\iota}\bm{b}\bm{\varsigma}}\bm{n}\bm{\varsigma}e^{-\hat{\iota}\bm{b}\bm{\varsigma}}}_{\bm{S}\bm{\varsigma}}
\sin \frac{\varphi}{2}=\cos \frac{\varphi}{2}+\bm{S}\bm{\varsigma}\sin \frac{\varphi}{2}=e^{\frac{\varphi}{2}\bm{S}\bm{\varsigma}}.
\end{gather}
As we can see, exponent $e^{\frac{\varphi}{2}\bm{S}\bm{\varsigma}}$ in (137) results in bispinor transformation  $e^{\hat{\iota}\bm{b}\bm{\varsigma}}\psi_{\pm}$, 
followed by multiplication by the unitary multiplier $e^{\mp i\frac{\varphi}{2}}$ only. Infinitesimal operator $\frac{\varphi}{2}\bm{S}\bm{\varsigma}$ of this transformation within the multiplier coincide with operator $\frac{\hbar}{2}i\bm{S}\bm{\varsigma}$.  For this operator, bispinor $e^{\hat{\iota}\bm{b}\bm{\varsigma}}\psi_{\pm}$ is a proper bispinor.

Transformation (137) with the exponent $e^{\frac{\varphi}{2}\bm{S}\bm{\varsigma}}$ is neither a rotation nor boost because
 $\bm{S}$ is an $\hat{\iota}$-complex number. 

Besides the property of the transformation with the exponent $e^{\frac{\varphi}{2}\bm{S}\bm{\varsigma}}$ discussed above, 
it also does not change the linear momentum of the fermion represented by this bispinor:  
\begin{gather}
e^{\frac{\varphi}{2}\bm{S}\bm{\varsigma}}p_\alpha\gamma^\alpha e^{-\frac{\varphi}{2}\bm{S}\bm{\varsigma}}=p_\alpha\gamma^\alpha .
\end{gather}
This is obvious because  in the reference frame $S_{-\bm{v}}$, (139) represents spatial rotation  $e^{\frac{\varphi}{2}\bm{n}\bm{\varsigma}}p_0\gamma^0 e^{-\frac{\varphi}{2}\bm{n}\bm{\varsigma}}$ of the 
4D linear momentum of the fermion (in its own proper reference frame).  This rotation does not change the 4D linear momentum of the fermion at rest:
\begin{gather}
e^{\frac{\varphi}{2}\bm{n}\bm{\varsigma}}p_0\gamma^0 e^{-\frac{\varphi}{2}\bm{n}\bm{\varsigma}}=p_0\gamma^0 .
\end{gather}
Transformation to the reference frame $S_{-\bm{v}}$
\begin{gather}
\underbrace{e^{\hat{\iota}\bm{b}\bm{\varsigma}}e^{\frac{\varphi}{2}\bm{n}\bm{\varsigma}}e^{-\hat{\iota}\bm{b}\bm{\varsigma}}}_
{e^{\frac{\varphi}{2}\bm{S}\bm{\varsigma}}}
\underbrace{e^{\hat{\iota}\bm{b}\bm{\varsigma}}p_0\gamma^0 e^{-\hat{\iota}\bm{b}\bm{\varsigma}}}_
{p_\alpha\gamma^\alpha}
\underbrace{e^{\hat{\iota}\bm{b}\bm{\varsigma}}e^{-\frac{\varphi}{2}\bm{n}\bm{\varsigma}}e^{-\hat{\iota}\bm{b}\bm{\varsigma}}}_
{e^{-\frac{\varphi}{2}\bm{S}\bm{\varsigma}}}=\nonumber\\=
\underbrace{e^{\hat{\iota}\bm{b}\bm{\varsigma}}p_0\gamma^0e^{-\hat{\iota}\bm{b}\bm{\varsigma}}}_{p_\alpha\gamma^\alpha} .
\end{gather}
leads to (139). Here
$p_\alpha\gamma^\alpha=e^{\hat{\iota}\bm{b}\bm{\varsigma}}p_0\gamma^0e^{-\hat{\iota}\bm{b}\bm{\varsigma}}$ ---
4D linear momentum $p_\alpha\gamma^\alpha=(mc\gamma^0-m\bm{v}\bm{\gamma})/\sqrt{1-\frac{v^2}{c^2}}$ in the
reference frame $S_{-\bm{v}}$.

Obviously, not only does spatial rotation $e^{\frac{\varphi}{2}\bm{n}\bm{\varsigma}}$ about the $\bm{n}$ axis on an angle $\varphi$ not
change the 4D linear momentum of a fermion at rest, but neither does rotation $e^{\bm{x}\bm{\varsigma}}$ about any axis $\frac{\bm{x}}{x}$ on any angle  $x$. Correspondingly, not only does transformation (139)  not change the linear momentum of the moving fermion, but  neither does any transformation $e^{\bm{S}(\bm{x})\bm{\varsigma}}p_\alpha\gamma^\alpha e^{-\bm{S}(\bm{x})\bm{\varsigma}}$ which contains an arbitrary 3D
pseudo-vector $\bm{x}$ instead of the Pauli-Lubanski pseudo-vector $\bm{n}$ (see expression (121) for $\bm{S}(\bm{x})\bm{\varsigma}$):
\begin{gather}
\bm{S}(\bm{x})\bm{\varsigma}=e^{\hat{\iota}\bm{b}\bm{\varsigma}_2}\bm{x}\bm{\varsigma}e^{-\hat{\iota}\bm{b}\bm{\varsigma}}=
\nonumber\\=
\frac{1}{mc}\Bigl(p_0\bm{x}-\frac{\bm{x}\cdot\bm{p}}{p_0+mc}\bm{p}+\hat{\iota}\bm{x}\times\bm{p}\Bigr)\bm{\varsigma}=
\nonumber\\=\frac{1}{mc}\bigl[(p_0\bm{y}-y_0\bm{p})+\hat{\iota}\bm{y}\times\bm{p}\bigr]\bm{\varsigma},
\end{gather}
where
\begin{gather}
y_0=\frac{\bm{x}\cdot\bm{p}}{mc},\hspace{7mm}\bm{y}=\bm{x}+\frac{\bm{x}\cdot\bm{p}}{mc}\frac{\bm{p}}{p_0+mc}.
\end{gather}
We can use simple calculations in order to prove equation (139) is true  when using operator $e^{\pm\bm{S(\bm{x})}\bm{\varsigma}}$:
\begin{gather}
e^{\bm{S(\bm{x})}\bm{\varsigma}}p_\alpha\gamma^\alpha e^{-\bm{S}(\bm{x})\bm{\varsigma}}=\nonumber\\=
(\cos x+\bm{S}\bm{\varsigma}\frac{\sin x}{x})p_\alpha\gamma^\alpha(\cos x-\bm{S}\bm{\varsigma}\frac{\sin x}{x})=
p_\alpha\gamma^\alpha.
\end{gather}
Because spatial rotations form the three-parameter subgroup of the Lorentz group, transformations  represented by exponents $e^{\bm{S}(\bm{x})\bm{\varsigma}}$ also 
form the three-parameter subgroup of the Lorentz group with the parameter $\bm{x}$.  It is the little Wigner's group \cite{w}, which belongs to $p_\alpha$ linear momentum.
Any transformation of this group does not change linear momentum $p_\alpha$.  Those transformations, for which $\bm{x}=\bm{n}\frac{\varphi}{2}$,
not only do not change linear momentum, but also change bispinors in a special way.  They  multiply them by unitary $e^{\mp i\frac{\varphi}{2}}$ only.

\section{Summary}
Lorentz transformations are important for all branches of physics because they reflects the structure of spacetime.  Thus,  the method of easy implementation of these 
transformations is also important. Lorentz's transformations are 4D rotations in spacetime. That is why this easy implementation must be similar 
to the rotations in in 3D  and 2D spaces.   \\  
The way to get this easy implementation is to represent the rotation operator as an exponent with an imaginary power: 
a)  in the 3D case, as an exponent with a vector-quaternion power  
b) in 4D space-time as an exponent with a hypercomplex power, which is the antisymmetric tensor of the second order in a hypercomplex representation.  \\  
Hypercomplex numbers are not commutative. So, even finding the exponent power, which is the product of the two exponents, becomes a complex problem.    
The rule of finding such a power for the Dirac numbers in an explicit form is presented in this paper (expressions (80), (85), (90)).  \\ 
This rule significantly simplifies consideration of the problems, which require the combination of several  Lorentz's transformations.  
The effectiveness of this rule is demonstrated for several problems, connected to Lorentz's transformations.    

\vspace{5mm}
\section{Acknowledgements}
The authors would like to thank Prof. Lev B.I., Prof. Lukyanets S.P.
and Prof. Denis Kovalenko for stimulating discussions.

\end{document}